\PassOptionsToPackage{hyphens}{url}
\documentclass[camera,letterpaper,nomarginnotes,nonarrowgutter]{jpaper}

\def\BibTeX{{\rm B\kern-.05em{\sc i\kern-.025em b}\kern-.08em
    T\kern-.1667em\lower.7ex\hbox{E}\kern-.125emX}}

\usepackage{fancyhdr}
\usepackage[normalem]{ulem}
\usepackage{amsmath}
\usepackage{mathtools}
\usepackage{tabularray}
\usepackage{xcolor,soul}
\usepackage[numbers,sort&compress]{natbib}
\usepackage{amssymb,amsfonts}
\usepackage{algorithmic}
\usepackage{graphicx}
\usepackage{textcomp}
\usepackage{pifont}
\usepackage[colorlinks=true,urlcolor=blue,citecolor=blue,linkcolor=blue]{hyperref}

\newcommand\VMcharacterization{~\cite{vm2,karakostas_characterIISWC,
isca2010-barr-trancache,5-levelpaging,contiguitas2023,radiantISMM21,bhattacharjeePACT2009,devirtualizingASPLOS2018,hash_dont_cache,virtualizationimplication,vm29,
vmhpcIISWC2024nick,kanellopoulos2025virtuoso,kanellopoulos2026cutting}}

\newcommand\VMlargepages{~\cite{park2020perforated,guvenilir2020tailored,ingensOSDI2016,talluriISCA1992,panwar2018making,panwar2019hawkeye,tridentMICRO2021,pham2015,mosaic2017MICRO,promotionHPCA2001,shadowpageISCA1998,duHPCA2015,vm42,vm43,partialMICRO2020,ganapathy98}}

\newcommand\VMcontiguity{~\cite{translationranger2019,flataAsplos2022,karakostas2015,chloe2020,hybridtlbISCA2017,flexpointerTACO2023,contiguitas2023,vm6,vm2,dmtASPLOS2024,margaritov2019prefetched}}

\newcommand\VMtlblthree{~\cite{sharedl3tlbISCA2011,distlltlbMICRO2018,gpustealing}}

\newcommand\VMtlbprefetching{~\cite{vavouliotis2021,morriganMICRO2021,kandiraju2002going,saulsbury2000recency,Bala1994SoftwarePA}}

\newcommand\VMtlbopts{~\cite{papadopoulou2015,latr,juan97,onlinesuperpagepromotionISCA1995,compilerdtlbISPASS2004,vm39,wood1986,skewedTLB,fan05}}

\newcommand\VMtlbreplacementpolicy{~\cite{deadTLBHPCA2021,chirpMICRO2020}}

\newcommand\VMtlball{~\cite{chirpMICRO2020,papadopoulou2015,latr,juan97,onlinesuperpagepromotionISCA1995,compilerdtlbISPASS2004,vm39,wood1986,skewedTLB,vavouliotis2021,morriganMICRO2021,margaritov2019prefetched,kandiraju2002going,saulsbury2000recency,Bala1994SoftwarePA,isca2010-barr-trancache,vm10,sharedl3tlbISCA2011,distlltlbMICRO2018,kanellopoulosMICRO2023victima}}

\newcommand\VMsoftwareTLB{~\cite{pomtlbISCA2017,csaltMICRO2017,softwareTLBNAS2013,softwareTLBISCA2013,bruceMMU1998,softcontrolcachesISCA1986,Nagle1993DesignTF,Bala1994SoftwarePA}}

\newcommand\VMtlbincache{~\cite{kanellopoulosMICRO2023victima,gputlbreachMICRO2021,ducati}}

\newcommand\AllUVM{~\cite{vogelTPDS2017,vogelTC2019,vogelPhD2018,kurthICCD2018,sonmezFPL2024,richterICCAD2022,cooperICS2024,kalkhofFPT2021,parkDATE2023uvmmu,choiTC2023,wangoasis,wanggrit,suvMICRO24,shinARIADNE2026,parkMICRO2025chiplet,haLATPC2025,fengHELIOSTAT2025,jangSOFTWALKER2025,jangPACT2024,fengBARRECHORD2024,liSTAR2024,kimMOST2025}}

\newcommand\HybridNVM{~\cite{yoon2012,banshee,li2017utility,zhao2014firm,salkhordeh2019analytical,meza2012enabling,ramos11,wangPantheraPLDI2019,chenHolisticTOCS2021,oliveiraOptaneAccess2023,nadigHarmonia2025,songPCMISMM2020,kokolisPageSeerHPCA2019,olsonHetMemTACO2022,wenMobileHybridTCAD2020, janus, daxvm}}

\newcommand\DisaggregatedMemory{~\cite{pberry2019,Gao2016,Shan2018,Korolija2021,Wang2020,Zuo2021,Maruf2020,Lim2009,Zhang2020,Yan2019,Angel2020,Lim2012,Peng2020,Bindschaedler2020,Katrinis2016,Aguilera2017,Aguilera2018,Rao2016,Calciu2021,Adya2019,Lagar2019,Pinto2020,Gu2017,Buragohain2017,Zervas2018}}
\usepackage{xspace}
\usepackage{color}
\usepackage{listings}
\usepackage{xcolor}

\definecolor{darkspringgreen}{rgb}{0.09, 0.45, 0.27}
\definecolor{denim}{rgb}{0.08, 0.38, 0.74}
\definecolor{darkolivegreen}{rgb}{0.33, 0.42, 0.18}
\definecolor{tangerine}{rgb}{0.95, 0.52, 0.0}
\definecolor{mahogany}{rgb}{0.75, 0.25, 0.0}

\newcommand{\pospt}{\textcolor{ForestGreen}{\raisebox{0.15ex}{\scriptsize$\blacktriangle$}}}
\newcommand{\midpt}{\textcolor{denim}{\raisebox{0.05ex}{\scriptsize$\bullet$}}}
\newcommand{\negpt}{\textcolor{mahogany}{\raisebox{0.15ex}{\scriptsize$\blacktriangledown$}}}
\newcommand{\posorangecell}[1]{%
  \textcolor{orange}{$\blacktriangle$}\ #1%
}
\newcommand{\poscell}[1]{\pospt\,#1}
\newcommand{\midcell}[1]{\midpt\,#1}
\newcommand{\negcell}[1]{\negpt\,#1}

\definecolor{seagreen}{rgb}{0.18, 0.55, 0.34}

\definecolor{profblue}{RGB}{0,90,200}
\newcommand{\profrev}[1]{#1}
\newcommand{\profreva}[1]{{\color{black}{#1}}}
\newcommand{\profrevc}[1]{{\color{black}{#1}}}

\usepackage{tikz}

\newcommand{\squishlist}{
 \begin{list}{$\circ$}
  { \setlength{\itemsep}{0pt}
     \setlength{\parsep}{0pt}
     \setlength{\topsep}{3pt}
     \setlength{\partopsep}{0pt}
     \setlength{\leftmargin}{1em}
     \setlength{\labelwidth}{1em}
     \setlength{\labelsep}{0.5em} } }

\newcommand{\squishend}{
  \end{list}  }

\newcommand*\circled[1]{\tikz[baseline=(char.base)]{\node[shape=circle,fill,inner sep=0.5pt] (char) {\textcolor{white}{#1}};}}
\definecolor{denim}{rgb}{0.08, 0.38, 0.74}
\definecolor{azure(colorwheel)}{rgb}{0.0, 0.5, 1.0}
\definecolor{greenp}{rgb}{0.0, 0.65, 0.50}
\definecolor{peach}{rgb}{0.97, 0.51, 0.47}
\definecolor{darkmagenta}{rgb}{0.55, 0.0, 0.55}
\definecolor{royalblue(web)}{rgb}{0.25, 0.41, 0.88}
\definecolor{ao(english)}{rgb}{0.0, 0.5, 0.0}
\definecolor{ForestGreen}{RGB}{34,139,34}

\usepackage[colorinlistoftodos,prependcaption,textsize=scriptsize]{todonotes}

\newcommand\todoFill[1]{}
\newcommand\discuss[1]{}
\newcommand\discussCritical[1]{}
\newcommand\fixSoon[1]{}

\newcommand\add[1]{}

\usepackage[bottom]{footmisc}
\usepackage{dblfloatfix}
\usepackage{array}
 \usepackage{booktabs}

\definecolor{ufogreen}{rgb}{0.1, 0.6, 0.4}
\definecolor{ufogreen}{rgb}{0.1, 0.6, 0.4}

\definecolor{hsviolet}{RGB}{138, 43, 226}
\newcommand{\hs}[1]{{\color{black}#1}}

\makeatletter
\newcommand{\drafttime}{%
  \@tempcnta=\time \divide\@tempcnta by 60
  \the\@tempcnta:%
  \@tempcntb=\@tempcnta \multiply\@tempcntb by 60
  \@tempcnta=\time \advance\@tempcnta by -\@tempcntb
  \ifnum\@tempcnta<10 0\fi
  \the\@tempcnta}
\makeatother

\newcommand{\system}{Revelator\xspace}

\usepackage{listings}
\usepackage{xcolor}
\usepackage[T1]{fontenc} 
\usepackage{zi4} 
\usepackage{caption} 
\usepackage{float} 

\lstdefinestyle{custompseudocode}{
  belowcaptionskip=1\baselineskip,
  breaklines=true,
  xleftmargin=\parindent,
  language=Python,
  showstringspaces=false,
  basicstyle=\small\ttfamily,
  keywordstyle=\bfseries\color{green!40!black},
  commentstyle=\itshape\color{purple},
  stringstyle=\color{orange},
  numbers=left,
  numberstyle=\scriptsize\color{black},
  numbersep=8pt,
  morekeywords={function}, 
  keywordstyle=[2]\bfseries, 
  escapeinside={*@}{@*}, 
  xleftmargin=2em, xrightmargin=0em, 
}

\definecolor{watermelon}{RGB}{232, 85, 74}

\definecolor{darkspringgreen}{rgb}{0.09, 0.45, 0.27}
\definecolor{denim}{rgb}{0.08, 0.38, 0.74}
\definecolor{darkolivegreen}{rgb}{0.33, 0.42, 0.18}
\definecolor{tangerine}{rgb}{0.95, 0.52, 0.0}
\definecolor{mahogany}{rgb}{0.75, 0.25, 0.0}
\definecolor{azure(colorwheel)}{rgb}{0.0, 0.5, 1.0}
\definecolor{greenp}{rgb}{0.0, 0.65, 0.50}
\definecolor{peach}{rgb}{0.97, 0.51, 0.47}
\definecolor{deepred}{RGB}{139, 0, 0}
\definecolor{darkmagenta}{rgb}{0.55, 0.0, 0.55}
\definecolor{royalblue(web)}{rgb}{0.25, 0.41, 0.88}
\definecolor{ao(english)}{rgb}{0.0, 0.5, 0.0}
\definecolor{ForestGreen}{RGB}{34,139,34}

\colorlet{ra}{darkspringgreen}
\colorlet{rb}{darkmagenta}
\colorlet{rc}{deepred}
\colorlet{rd}{tangerine}
\colorlet{re}{mahogany}
\colorlet{rf}{darkolivegreen}
\colorlet{cq}{denim}

\newcommand{\revF}[1]{#1}
\newcommand{\revCQ}[1]{#1}

\newcommand\revmid[2]{}

\newcommand{\revahigh}[1]{#1}
\newcommand{\revbhigh}[1]{#1}
\newcommand{\revchigh}[1]{#1}
\newcommand{\revdhigh}[1]{#1}
\newcommand{\revehigh}[1]{#1}
\newcommand{\revfhigh}[1]{#1}
\newcommand{\revcqhigh}[1]{#1}

\AtBeginDocument{%
  \providecommand\BibTeX{{%
    Bib\TeX}}}

\title{Revelator: Rapid Data Fetching \profreva{via System-Software-Guided} Hash-based Speculative Address Translation}

\author{
    Konstantinos Kanellopoulos$^1$ \hspace{0.5em}
    Konstantinos Sgouras$^1$ \hspace{0.5em}
    Harsh Songara$^1$ \vspace{0.2em} \\
    Andreas Kosmas Kakolyris$^1$ \hspace{0.5em}
    Vlad-Petru Nitu$^1$ \hspace{0.5em}
    Spiros Galanopoulos$^1$ \hspace{0.5em}
    Rahul Bera$^1$ \vspace{0.2em} \\
    Konstantina Koliogeorgi$^1$ \hspace{0.5em}
    Rakesh Kumar$^2$ \hspace{0.5em}
    Onur Mutlu$^1$ \vspace{0.5em} \\
    \normalsize{
        $^1$ETH Z\"urich \hspace{0.5em} $^2$NTNU
    }
}

\begin{document}

\maketitle

\thispagestyle{plain}
\pagestyle{plain}

\begin{abstract}

    Address translation is a significant performance bottleneck in modern computing systems.
    Predicting the physical address (PA) of the requested data before address translation completes is a promising technique to \profrev{hide the latency of address translation}.
    However, accurately predicting the PA based on the virtual address (VA) is highly challenging due to the inherent unpredictability of the VA-to-PA mappings that conventional operating systems introduce.
    Prior works try to introduce predictability into VA-to-PA mappings but exhibit two key shortcomings: (i) they rely on the availability of large pages or VA-to-PA contiguity, and (ii) they \profrev{store speculation-related metadata in costly hardware components whose power \profreva{and} area overheads may not always justify their benefits}.

    We introduce Revelator, a new hardware-OS cooperative \profrev{technique} that uses hashing to enable highly accurate speculative address translation with \profreva{small} system modifications.
    At the OS level, Revelator employs a tiered hash-based memory allocation policy for \emph{both} program data and last-level page table entries (PTEs) to establish predictable VA-to-PA and VA-to-PTE mappings. 
    At the hardware level, after an L2 TLB miss, a lightweight speculation engine uses the OS's hash functions to predict the VA-to-PA and VA-to-PTE mappings and prefetch the corresponding \profrev{cache blocks} before address translation completes, \profrev{thereby} \emph{both} hiding address translation latency \emph{and} accelerating page table walks (PTWs).
    Revelator comes with two key benefits: (i) \profreva{it does not rely on large pages or VA-to-PA contiguity} to achieve high speculation accuracy,
    and (ii) it requires \profrev{small} OS and hardware modifications.
    
    \profrev{Our evaluation across 11 data-intensive workloads shows that, (i) in single-core systems\profreva{,} Revelator \profreva{provides} an average speedup of 15.3\% over the state-of-the-art speculative address translation technique under high memory fragmentation, and (ii) in virtualized environments, by accurately predicting \emph{both} guest and host physical addresses, 
    Revelator \profreva{provides} a 13.6\% average speedup over Nested Paging. In multicore systems, Revelator's benefits scale with core count, \profreva{enabling a} speedup of \profrev{1.40$\times$} (1.50$\times$) over Transparent Huge Pages (THP) across 30 \profreva{server workload mixes from Google} in a 16-core system under medium (high) memory fragmentation.}
     These gains come at very small hardware cost. Our RTL synthesis shows that Revelator incurs only 0.02\% area and 0.03\% power overheads on top of a high-end server-grade CPU. Revelator is freely available at
    \href{https://github.com/CMU-SAFARI/Virtuoso}{github.com/CMU-SAFARI/Virtuoso}.
\end{abstract}

\section{Introduction}
\label{sec:introduction}

As modern applications become increasingly data-intensive, address translation has emerged as a major performance bottleneck\profreva{\VMcharacterization}.
 Recent studies from academia and industry demonstrate that address translation can account for up to 40-45\% of total execution time (15-20\% with large pages)~\cite{radiantISMM21,contiguitas2023,karakostas_characterIISWC,vm2,warehouse,kanellopoulos2026cutting,dmtASPLOS2024,kanellopoulos2023utopia,kanellopoulosMICRO2023victima,elastic-cuckoo-asplos20,karakostas2015,chloe2020,vbi,midgard}.
 Address translation overheads are projected to grow as architectures transition toward larger physical address spaces, driven by paradigms like disaggregated memory\DisaggregatedMemory, tiered memory systems\HybridNVM, and unified virtual memory (UVM) in heterogeneous architectures\AllUVM.

\textbf{Speculative Address Translation.} \profreva{Speculation is a} promising technique to mitigate address translation overheads. 
By predicting the physical address (PA) corresponding to a virtual address (VA) before resolving the VA-to-PA mapping, the processor can begin fetching data \profreva{using the PA} immediately, without waiting for address translation to complete. 
This allows the speculative memory \profrevc{data} access to overlap with the ongoing translation lookaside buffer (TLB) \profreva{lookup} and page table walk, reducing or hiding address translation latency. 
However, accurately predicting the PA from the VA remains highly challenging due to the inherent unpredictability of the VA-to-PA mappings that conventional \profreva{OSes} introduce.
Because conventional OSes flexibly map virtual pages to \profreva{\emph{any}} available physical frame, the exact mapping depends heavily on dynamic system conditions such as (i) memory fragmentation and (ii) \profreva{memory} allocation history.

\textbf{Shortcomings of Prior Works.} Prior works on speculative address translation~\cite{spectlbISCA2011,spec-gpuMICRO2024,chloe2020,pham2015} attempt to form predictable VA-to-PA mappings. However, they rely on strong assumptions that limit their applicability and effectiveness in real systems. 
First, they rely on the availability of free large pages (e.g., 2MB) or VA-to-PA contiguity, neither of which can be guaranteed in practice, especially in  systems \profrevc{with highly-fragmented memory} (e.g., datacenters) or when multiple applications compete for memory~\cite{cbmm,mansi2024characterizingphysicalmemoryfragmentation}.
Second, they require significant changes to the processor architecture, including new power- and area-hungry hardware components that store speculation-related metadata. For example, Barr et al.~\cite{spectlbISCA2011} propose SpecTLB, a hardware TLB that opportunistically stores VA-to-PA mappings for reserved but not-yet-promoted large pages. 
However, as shown in \S\ref{sec:spectlbmotivation}, \profrevc{such schemes provide low benefits even under modest memory fragmentation.}

    \textbf{Our goal} is to design a speculation mechanism that hides address translation latency in modern systems without relying on large pages, VA-to-PA contiguity, or costly hardware structures to store speculation metadata.
    We introduce \emph{Revelator}, a new hardware-OS cooperative scheme that uses hashing to enable highly accurate speculative address translation with \profrevc{small} system modifications.
    \textbf{The key idea} of Revelator is to have the OS establish a \emph{predictable} hash-based virtual-to-physical mapping, which the hardware exploits to perform highly accurate speculati\profrevc{ve address translation to predict which data to fetch.}
    The key benefits of Revelator are twofold:
    (1) \profreva{its prediction accuracy} \profrev{is not affected by} memory fragmentation, as \profreva{\emph{per-page}} hash-based allocation succeeds with a probability that depends \emph{only} on memory utilization and \emph{not} on how free frames are distributed (e.g., available free large pages, VA-to-PA contiguity), and
    (2) it requires \profreva{small} OS and hardware modifications, i.e., minor extensions to the buddy allocator to enable hash-based data allocation and a lightweight speculation engine that generates candidate PAs based on the same hash functions used by the OS.

    \textbf{Key Mechanism.} \profreva{At the OS level, Revelator employs a \emph{tiered} hash-based allocation policy to increase the likelihood of establishing a predictable VA-to-PA mapping. When allocating a physical frame for a virtual page, the OS performs up to $N$ hash-based allocation attempts (i.e., allocation tiers). In each attempt $i$, the OS applies the same hash function to the virtual page number ($VPN$) with a tier-specific seed, yielding a candidate physical page number ($PPN_i$). The OS allocates the first $PPN_i$ that corresponds to a free physical frame, falling back to the conventional allocator only if all $N$ candidates are \profreva{already allocated}. Because each hash-based allocation attempt acts as an independent trial, the probability that all $N$ candidates are already occupied decreases exponentially with $N$, ensuring a high success rate for establishing predictable VA-to-PA mappings even under high memory utilization.}

    \profreva{ At the hardware level, on an L2 TLB miss, a lightweight speculation engine first recomputes the same $N$ hash-derived candidates and combines each candidate $PPN_i$ with the page offset, producing $N$ candidate PAs. 
    Revelator does not blindly issue \profrevc{speculative accesses to all $N$ candidate PAs}. Instead, a speculation degree filter selects how many candidates to fetch based on the recent usefulness of each allocation tier and the current memory bandwidth pressure. 
    Thus, the speculation engine issues at most $N$ speculative data requests, and often fewer depending on system conditions.
    If one of the candidates is correct, the data can arrive before the PTW completes, effectively hiding address translation latency.}

Revelator applies the same hash-based allocation and speculation to the page table (PT) itself. \profreva{Similar to data page allocation}, the OS attempts to place each last-level PT frame at $PPN_i = Hash_i(VPN \gg 9)$, where the 9-bit shift accounts for the 512 VPNs that share each PT frame. On an L2 TLB miss, the hardware speculatively fetches this last-level PT frame in parallel with the earlier levels of the walk. Altogether, Revelator achieves two levels of concurrency: (1) it overlaps the final data fetch with the entire page table walk, and (2) it overlaps the fetch of the last-level PT with earlier stages of the walk itself.

\textbf{Integration with Large Pages and NUMA.} Revelator fully supports large-page allocation and NUMA-aware placement, two widely deployed features of modern OSes. For large pages, Revelator+THP combines the TLB-reach benefits of Transparent Huge Pages (THP)~\cite{corbet2011,corbet2017} when 2MB pages are available with Revelator's hash-based 4KB speculation when fragmentation makes large pages unavailable, \profreva{enabling} high performance across the fragmentation spectrum (\S\ref{sec:revelator-thp}). For NUMA systems, Revelator preserves the OS's node-selection policy and introduces a new lightweight, Bloom-filter-based data structure that tracks which NUMA node each virtual region maps to, enabling the hardware to speculate accurately even when a thread accesses pages on remote NUMA nodes (\S\ref{sec:NUMA}).

\textbf{Evaluation Methodology.} 
We prototyped Revelator's tiered hash-based allocation \profreva{mechanism} in the Linux kernel (v.6.10.8)~\cite{linux-610} to demonstrate its feasibility and effectiveness.
We evaluate the effectiveness of Revelator in single-core, multi-core, and virtualized environments using \profreva{Virtuoso~\cite{kanellopoulos2025virtuoso,virtuosogithub}}, a validated simulation methodology that enables high-accuracy full-system simulation ported on top of \profreva{Sniper~\cite{sniper,snipergithub}}, an event-driven multi-core simulator. In single-core systems, we evaluate Revelator across 11 translation-intensive benchmarks from GraphBIG~\cite{Lifeng2015}, GUPS~\cite{Plimpton2006}, DLRM~\cite{dlmr}, GenomicsBench~\cite{genomicsbench}, and XSBench~\cite{Tramm2014}. In multi-core systems, we evaluate Revelator on 30 mixes from 140 Google Server Traces~\cite{google_workload_traces_v2_dynamorio}. We compare Revelator against \profreva{eight} state-of-the-art address translation schemes~\cite{reserve,chloe2020,corbet2011,spectlbISCA2011,mosaicpagesASPLOS2023,pomtlbISCA2017,elastic-cuckoo-asplos20,margaritov2019prefetched,dmtASPLOS2024} and a system with a very large 64K-entry L2 TLB.

\textbf{Key Results.} Our evaluation yields seven key results that show Revelator's effectiveness.
First, our Linux prototype, running on a real high-end system, shows that Revelator's tiered hash-based allocator using only \profreva{three} hash functions successfully allocates more than 70\% of pages even under high memory utilization \profreva{levels} (Fig.~\ref{fig:linux_kernel}), with minimal overhead (0.078\% on average) on total execution time.
Second, in single-core systems, Revelator \profreva{provides} a \profreva{19\%} (25\%) average speedup over THP, and an \profreva{8.5\%} (15.3\%) average speedup over SpOT~\cite{chloe2020}, the state-of-the-art speculative address translation technique, under medium (high) memory fragmentation.
 Revelator \profreva{provides} these gains while minimizing: (1) the number of speculative L2 \profreva{data cache fills} that lead to harmful evictions and (2) the bandwidth spent on unused speculative data fetches.
Third, in virtualized environments, Revelator ~\profreva{provides} a \profreva{13.6\%} average speedup over Nested Paging~\cite{amdnested} with THP~\cite{corbet2011} (NP+THP) by accurately predicting both guest and host physical addresses.
Fourth, in multi-core systems, Revelator's benefits scale with core count, \profreva{enabling} an average speedup of $1.40\times$ ($1.50\times$) over THP at 16~cores under medium (high) memory fragmentation, while sustaining 87--88\% speculation accuracy across all core counts.
Fifth, in an 8-node/64-core NUMA system, Revelator outperforms SpOT by up to $1.20\times$.
Sixth, Revelator+THP reduces energy consumption by 5.5\% over THP in a system with medium fragmentation, and by 2.2\% over Mosaic Pages~\cite{mosaicpagesASPLOS2023}, a scheme with very large TLB reach.
Seventh, our RTL synthesis results show that Revelator incurs only 0.02\% area and 0.03\% power overheads on top of a high-end server-grade CPU~\cite{cascadelake}.

In this work, we make the following contributions:

\begin{itemize}
\item We introduce \emph{Revelator}, a new hardware-OS cooperative scheme that uses \profreva{predictable} hashing to enable highly accurate speculative address translation with minimal system modifications. \textbf{The key idea} of Revelator is to have the OS establish a predictable hash-based virtual-to-physical mapping, which the hardware exploits to perform highly accurate speculation. \profrevc{This both hides and reduces address translation latency} and provides two key benefits: (i) \profrev{Revelator is not affected by} memory fragmentation and (ii) it requires \profreva{very small} OS and hardware modifications.
\item We prototype Revelator's allocator in the Linux kernel (v.6.10.8), evaluate Revelator via simulation across diverse benchmarks \profrevc{against nine prior techniques}, and validate its hardware costs via RTL synthesis. Our results demonstrate significant performance and energy gains in single-core, multi-core, NUMA, and virtualized systems \profrevc{compared to all prior techniques}. Revelator is freely available at \href{https://github.com/CMU-SAFARI/Virtuoso}{github.com/CMU-SAFARI/Virtuoso}.
\end{itemize}

\section{Motivation}
\label{sec:motivation}

\subsection{Speculative Address Translation}

\begin{figure}[h!]
    \centering
    \includegraphics[width=1.0\linewidth]{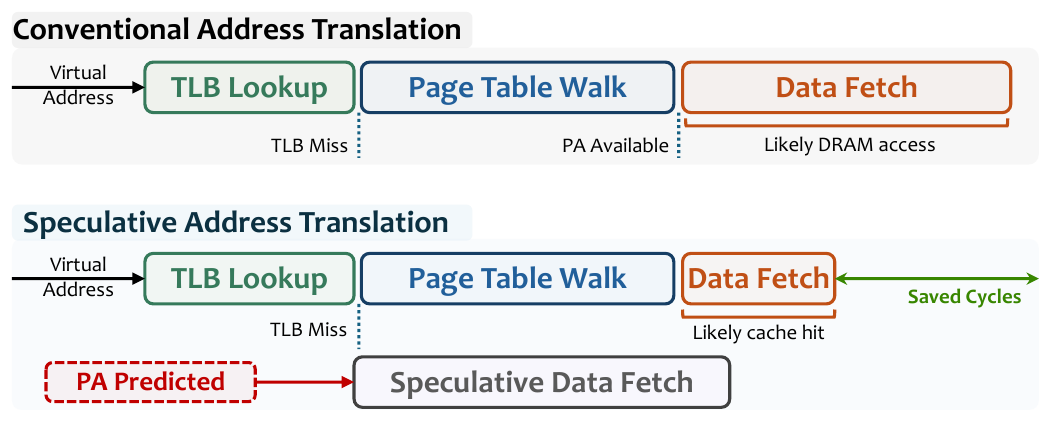}
    \caption{Data fetch sequence under (top) conventional address translation and (bottom) speculative address translation.}
    \label{fig:speculative_translation}
\end{figure}

Previous works~\cite{spectlbISCA2011,spec-gpuMICRO2024,chloe2020,flash-specACCESS2025,pham2015} hide address translation latency by speculating \profrevc{on} the physical address, allowing the processor to issue early memory accesses
and fetch data before address translation completes: a process called \emph{speculative address translation}.
Figure~\ref{fig:speculative_translation} shows the data fetch sequence under conventional and speculative \profrevc{address translation}.
In a conventional system, address translation and data fetching happen sequentially. The processor first performs a TLB lookup to translate the virtual address (VA) to a physical address (PA). If the TLB lookup results in a miss, the processor initiates a page table walk (PTW) to resolve the VA-to-PA mapping. Only after the PTW completes can the processor issue the memory access to fetch the data. This sequential process can lead to significant performance degradation, especially when the PTW requires high-latency DRAM accesses to complete. 
With speculative address translation, the processor predicts the physical address and issues an early speculative memory access \profrevc{to fetch the data} before starting the PTW. 
If the prediction is correct, the speculative fetch can partially or fully overlap data fetching with address translation, hiding the latency of address translation and improving performance.
If the prediction is incorrect, the processor fetches an irrelevant cache block. Such mispredictions do not affect correctness, but consume memory bandwidth, waste energy, and may pollute the cache.

\subsection{Challenges in Speculative Address Translation}

Predicting the physical location of data based on the virtual address is challenging because conventional OSes do not establish stable, easily predictable VA-to-PA mappings/\footnote{Predicting PAs using other program features (e.g., PCs), beyond VAs, is not meaningful because in conventional OSes the faulting VA and its properties (e.g., access type, permissions) are the only relevant program-visible features that affect PA selection.}
First, modern OSes use fully-associative virtual memory: during a page fault, the allocator maps the faulting virtual page to any currently available physical frame~\cite{osdev_page_frame_allocation}. Second, memory fragmentation and memory utilization continually change which frames are available. Together, these effects make VA-to-PA mappings highly dependent on runtime conditions, limiting 
the potential for accurate PA prediction.

\subsection{Limitations of Prior Works}
\label{sec:spectlbmotivation}
SpecTLB~\cite{spectlbISCA2011}, the first work that proposed speculative address translation, relies on the OS to reserve large pages in order to achieve accurate hardware-side PA prediction.  
On the OS side, a FreeBSD-like~\cite{reserve} OS allocation scheme reserves candidate 2MB regions but does not immediately use them as large pages. \profrevc{Instead, it allocates 4KB pages from the reserved region and only promotes the region to a large page when the OS detects that the region is highly utilized.} On the hardware side, SpecTLB adds a TLB structure that records reserved but not-yet-promoted large-page regions and predicts the PA by speculatively assuming a fixed VA-to-PA offset within each reserved region. 

SpOT~\cite{chloe2020} extends this idea from fixed-size large pages to arbitrarily-large contiguous virtual-to-physical mappings.
On the OS side, SpOT uses contiguity-aware paging (CA-paging) to sustain contiguous VA-to-PA mappings over time. Whenever possible, CA-paging places pages from the same virtually contiguous memory object, called a Virtual Memory Area (VMA) in Linux~\cite{vma}, in a contiguous physical segment, while preserving standard demand paging (i.e., pages are allocated on demand).
On the first fault to a VMA, CA-paging searches a contiguity map, an OS metadata structure that tracks available contiguous physical segments, to find a free physical segment that can accommodate the VMA.
It then allocates the faulting page in that segment, records the resulting VA$\rightarrow$PA offset for the VMA (i.e., the fixed PA-minus-VA displacement implied by that first placement), and uses that offset as a guide to allocate later-faulted pages from the same VMA in a best-effort manner.
This way, CA-paging tries to establish contiguous VA-to-PA mappings over time, which SpOT leverages to perform accurate speculative address translation.
On the hardware side, SpOT leverages the contiguous mappings created by CA-paging and caches them as contiguity descriptors in a small prediction table \profrevc{indexed by the load/store program counter (PC).}
Each descriptor records the virtual range (base and bounds), access permissions, and a speculative VA$\rightarrow$PA offset for a recently observed mapping.
On a TLB miss, the processor uses the load/store PC to retrieve a contiguity descriptor, checks whether the missed VA falls within the descriptor's range, and computes a speculative physical address as VA~+~offset.

\begin{table*}[!t]
\centering
\vspace{-1mm}
\caption{\revCQ{Comparison between (i) contiguity-based and (ii) Revelator's hash-based speculative address translation techniques.}}
\label{tab:contiguity-comparison}
\footnotesize
\begin{tblr}{
    width = \textwidth,
    colspec = {Q[l,wd=0.16\textwidth] Q[l,wd=0.24\textwidth] Q[l,wd=0.30\textwidth] Q[l,wd=0.25\textwidth]},
    column{1} = {font=\bfseries},
    row{1-2} = {font=\bfseries, halign=c},
    colsep = 3pt,
    rowsep = 1pt,
    vline{1} = {0.6pt},
    vline{2} = {1}{-}{0.6pt},
    vline{2} = {2}{-}{0.6pt},
    vline{3-4} = {0.35pt},
    vline{Z} = {0.6pt},
    hline{1} = {0.8pt},
    hline{3} = {1}{-}{0.6pt},
    hline{3} = {2}{-}{0.6pt},
    hline{4-9} = {0.35pt},
    hline{Z} = {0.8pt},
}
\SetCell[r=2]{m,c} \textit{Property} & Structured contiguity & Arbitrary contiguity & Hash-based allocation \\
 & (SpecTLB~\cite{spectlbISCA2011}) & (SpOT~\cite{chloe2020}) & (This work: Revelator) \\
Placement invariant & Fixed-size 2MB pages & Arbitrarily-large contiguous physical segments & Per-page hash-based placement\\
Speculation metadata & None & Contiguity descriptors per memory object & Counters per hash function \\
Hardware complexity & \negcell{Extra TLB for reserved large pages} & \negcell{Cache for contiguity descriptors} & \poscell{Hash function circuitry and counters} \\
Sensitivity to fragmentation & \negcell{Single "hole" prevents reserving a large page$^{*}$} & \posorangecell{Speculation coverage and accuracy drops as "holes" increase} & \poscell{Successful hash-based allocation depends only on memory utilization} \\
Allocation interference & \negcell{Allocation interference creates holes} & \negcell{Allocation interference disrupts contiguity} & \poscell{Independent of allocation interference} \\
Memory allocation & \poscell{Fewer physical memory allocations} & \midcell{Minimal impact on latency} & \negcell{Small increase in latency} \\
\end{tblr}
\vspace{0.5mm}
\begin{minipage}{\textwidth}
\footnotesize $^{*}$ A "hole" is an occupied page within an otherwise contiguous physical region that splits a large block into smaller non-contiguous segments, creating memory fragmentation.
\end{minipage}
\vspace{-1mm}
\end{table*}

\subsubsection{Limits of Contiguous Multi-Page Mappings}
Table~\ref{tab:contiguity-comparison} \profrevc{shows the key properties of
contiguity-based speculative address translation techniques that rely on (i) structured contiguity (e.g., SpecTLB~\cite{spectlbISCA2011}) or (ii) arbitrary contiguity (e.g., SpOT~\cite{chloe2020}).}
Both SpecTLB~\cite{spectlbISCA2011} and SpOT~\cite{chloe2020} depend on preserving \emph{multi-page invariants}: structured and fixed-size multi-page physical regions (e.g., $512\times$ 4KB pages) in SpecTLB~\cite{spectlbISCA2011} or arbitrarily-large (i.e., base and bounds of a single mapping can be arbitrary) contiguous physical segments in SpOT~\cite{chloe2020}. 
These invariants are challenging to maintain when (i) memory fragmentation, which is observed to be high in mobile and large-scale systems (e.g., datacenters)~\cite{mansi2024characterizingphysicalmemoryfragmentation,contiguitas2023}, severely limits the available physical memory contiguity, and 
(ii) co-running applications compete for contiguous allocations, \profrevc{a phenomenon we call \emph{allocation interference}}, which can rapidly increase memory fragmentation even when the system is lightly loaded.

We quantitatively analyze the sensitivity of \profrevc{contiguity-based speculative address translation techniques} to fragmentation and allocation interference, which motivates Revelator's hash-based placement invariant. We focus on SpOT~\cite{chloe2020} as it is the most recent work that relies on contiguity, but our analysis applies to any approach that relies on contiguous multi-page mappings for speculation (e.g., SpecTLB~\cite{spectlbISCA2011}).

\subsubsection{Sensitivity to Fragmentation}
\begin{figure}[h!]
        \centering
        \includegraphics[width=1.0\linewidth]{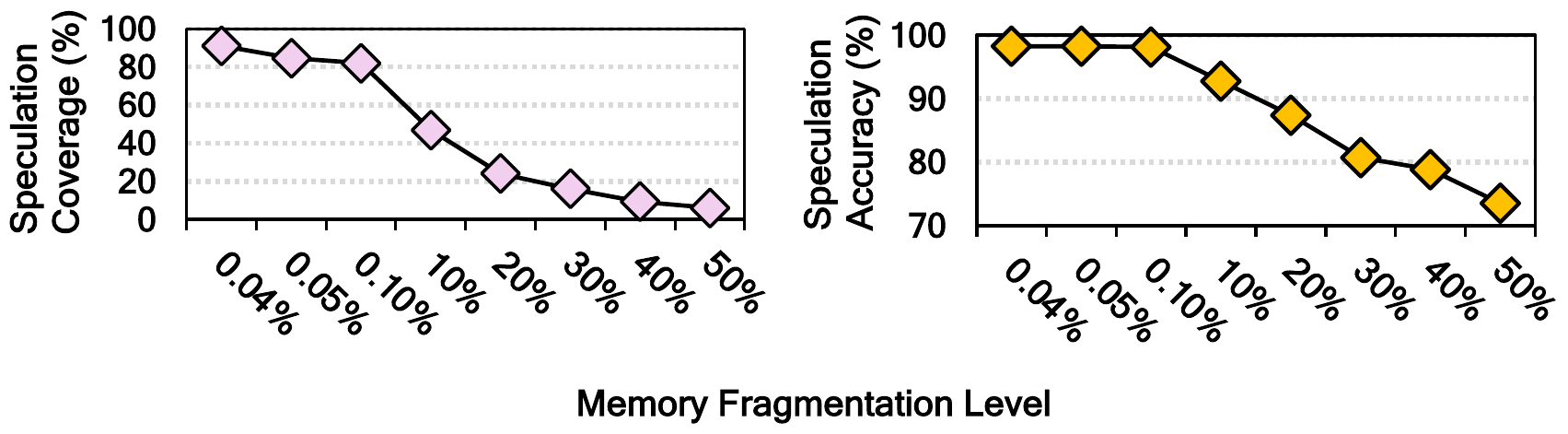}
        \vspace{-3mm}
        \caption{Speculation coverage (left) and accuracy (right) of SpOT~\cite{chloe2020}, a state-of-the-art  
        speculative address translation technique, in a system with different memory fragmentation levels.}
        \label{fig:spot_metrics}
\end{figure}

To quantify SpOT's sensitivity to fragmentation, Figure~\ref{fig:spot_metrics} shows the average speculation \profrevc{coverage (i.e., the fraction of TLB misses that are speculatively translated)} and accuracy \profrevc{(i.e., the fraction of correct speculative address translations)} across 11 translation-intensive workloads at different fragmentation levels (see \S\ref{sec:methodology} for details). 
We make three key observations. 
(1) Under low fragmentation ($<$1\%), SpOT delivers near-perfect accuracy ($\approx$100\%) and high coverage ($>$85\%). 
(2) At medium fragmentation levels (10--20\%), accuracy drops to $\approx$70\% and coverage degrades sharply: at 10\% fragmentation, coverage \profrevc{reduces} to $\approx$40\%, and beyond 20\% fragmentation, SpOT speculates on fewer than 20\% of TLB misses. 
(3) At high fragmentation ($\approx$50\%), SpOT provides very low coverage (below 10\%). 
Together, these observations show that the effectiveness of  contiguity-based \profrevc{speculative address translation techniques} (e.g., SpOT) heavily depends on memory fragmentation. In \S\ref{sec:results:comparison}, we further show the performance gains \profrevc{provided} by SpOT and Revelator under different fragmentation levels over the baseline system.

\subsubsection{Sensitivity to allocation interference}
SpOT's data placement strategy is effective when a small number of VMAs can be mapped to large contiguous physical segments, but it becomes fragile under allocation interference. Since SpOT does not preallocate the entire VMA, other VMAs or co-running applications can consume pages inside the targeted physical segment before subsequent page faults occur. Interleaved allocations create holes, disrupt the contiguity that SpOT relies on for speculative translation, and force the system to fall back to the conventional allocation policy (i.e., fully-associative data placement).

\begin{figure}[h!]
    \centering
    \hspace*{-3mm}
    \includegraphics[width=1.05\linewidth]{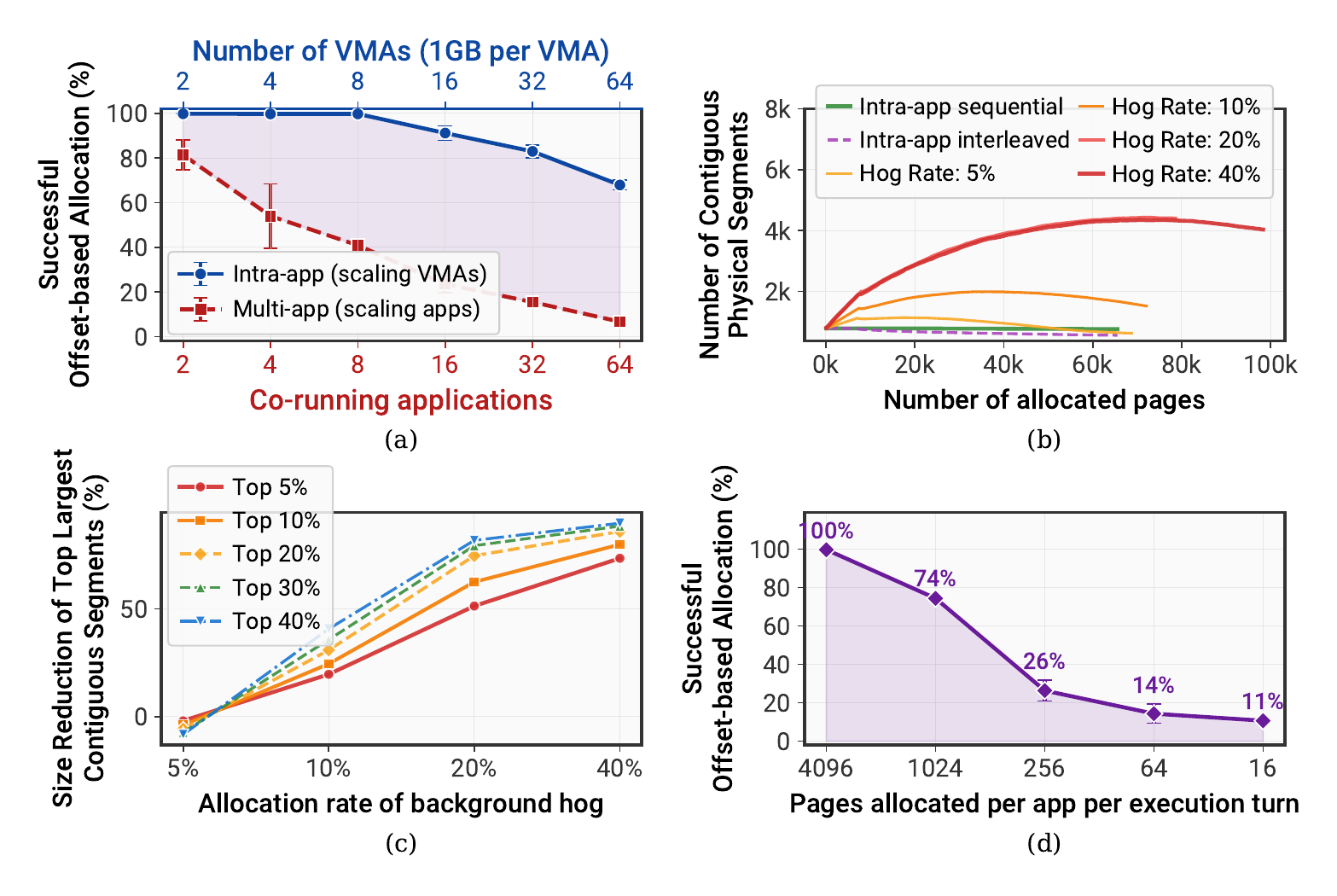}
    \caption{Sensitivity of SpOT to allocation interference: (a) ratio of successful offset-based allocations as the number of VMAs or co-running applications increases, (b) size reduction of the largest contiguous physical segments under background memory hogging, (c) increase in the number of contiguous physical segments under background memory hogging, and (d) successful offset-based allocations under finer-grained allocation interleaving.}
    \label{fig:eval:spot-multitenancy}
\end{figure}

Figure~\ref{fig:eval:spot-multitenancy} quantifies how allocation interference affects (i) the available physical memory contiguity  and (ii) the ratio of successful offset-based allocations, i.e., the fraction of allocations that follow the VMA's recorded VA$\rightarrow$PA offset. 
In this experiment, (1) memory is minimally fragmented (i.e., memory utilization is less than 1\%) to isolate the effects of allocation interference, (2) we use a microbenchmark that allocates memory for a configurable number of 1GB VMAs, and (3) we execute applications in a simulated system that employs a round-robin \profrevc{thread} scheduler and 128GB of main memory (see \S\ref{sec:methodology} for details about the simulation methodology).

Figure~\ref{fig:eval:spot-multitenancy} (a) shows the ratio of successful offset-based allocations (i) as the number of VMAs within one application increases and (ii) as the number of co-running applications increases. We make two key observations. First, increasing the number of VMAs within a single application gradually reduces the ratio of successful offset-based allocations. The success rate remains near 100\% with up to 8 VMAs and stays above 60\% even at 64 VMAs. Second, increasing the number of co-running applications reduces the ratio of successful offset-based allocations from ${\approx}$80\% with 2 applications to ${\approx}$5--10\% with 64 applications. We conclude that SpOT is more sensitive to inter-application interference than to the number of VMAs within a single application, because concurrent applications interleave independent allocation streams and create irregular allocation patterns that rapidly disrupt the available \profrevc{physical memory} contiguity.

Figure~\ref{fig:eval:spot-multitenancy} (b) shows the number of free contiguous physical segments for different hog rates as a foreground application runs alongside the background hog. We observe that as the hog rate increases and as the foreground application allocates more memory, the number of free contiguous physical segments grows from a few dozen to several thousand. This growth in the number of segments indicates that the free physical memory becomes highly fragmented, thereby reducing SpOT's effectiveness \profrevc{(as shown in \S\ref{sec:results:comparison})}.

Figure~\ref{fig:eval:spot-multitenancy} (c) measures how background memory hogging \profrevc{(i.e., allocation and freeing of pages performed by a co-running background process)} affects the largest available contiguous physical segments.
In this experiment, we run a background memory hog that allocates and frees pages at different rates (i.e., hog rate of $N$\% means that the background hog allocates $N$ pages per 100 pages allocated by the foreground application) alongside the foreground application. We make two key observations. First, even a low hog rate (20\%) reduces the largest contiguous physical segments by 50--80\%. Second, as the hog rate increases, the largest contiguous physical segments degrade further, and at 80\% hog rate, the largest contiguous segment consists of only a few pages ($\approx$32). We conclude that even moderate background allocation activity can break the large free contiguous physical segments into substantially smaller ones, which in turn significantly reduces SpOT's speculation coverage and accuracy \profrevc{(as shown in \S\ref{sec:results:comparison})}.

Figure~\ref{fig:eval:spot-multitenancy} (d) shows the effect of allocation interleaving granularity with 16 applications running concurrently. We vary the allocation turn size (i.e., the number of pages each application allocates before the OS scheduler switches execution to the next application) from 4096 pages down to 16 pages.
We observe that reducing the allocation turn size from 4096 pages to 16 pages reduces successful offset-based allocation from 100\% to 11\%. A smaller allocation turn size interleaves allocation requests from different applications more frequently, creating holes inside the physical segments that SpOT tries to preserve. We conclude that SpOT's offset-based placement breaks down rapidly under fine-grained allocation interleaving.

Taken together, these results show that SpOT's reliance on contiguous physical regions makes speculation fragile under realistic allocation dynamics: competing VMAs, co-running applications, and background allocation activity all break large free regions into smaller pieces, reducing offset-based allocation success. \profrevc{Our goal} in this work is to design a speculative address translation mechanism that does \emph{not} rely on contiguous multi-page mappings and can instead provide high speculation coverage and accuracy even under high fragmentation and \profrevc{high loads of} allocation interference.

\section{\system: Design Overview}
\label{design_overview}

\begin{figure}[h!]
    \centering
    \includegraphics[width=1.0\linewidth]{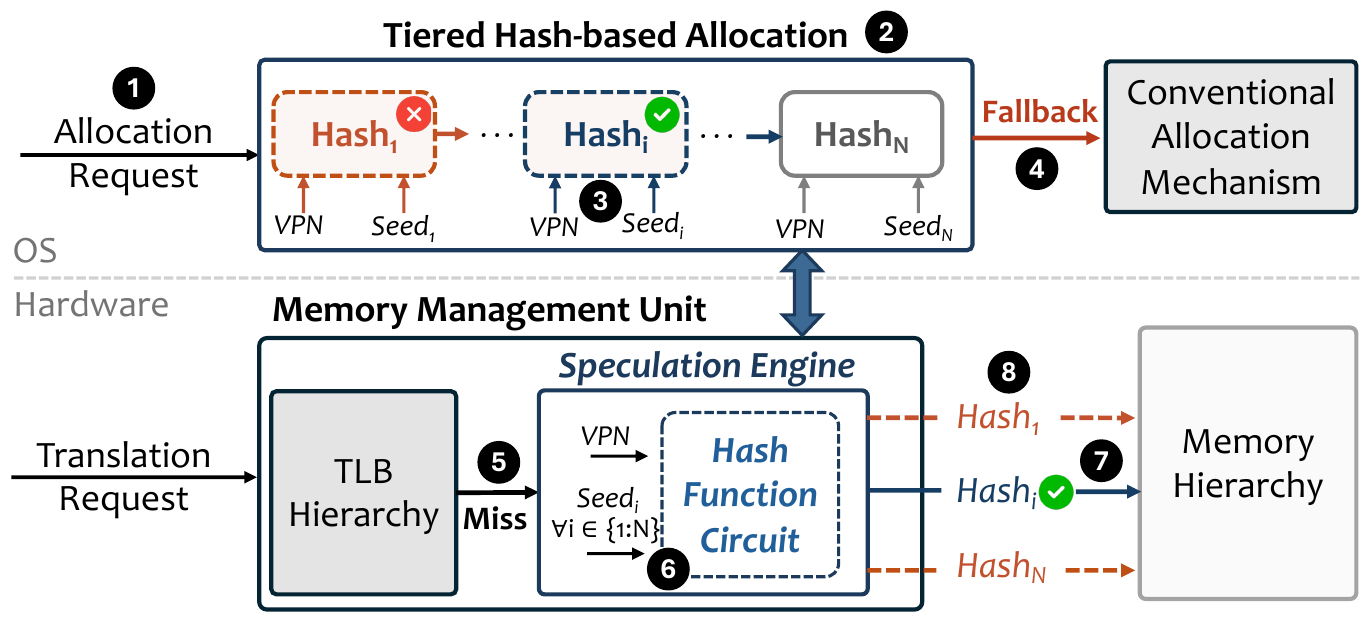}
    \vspace{-3mm}
    \caption{Revelator Overview}
    \label{fig:overview}
\end{figure}

We present Revelator, a hardware-OS cooperative scheme that uses hashing to enable highly accurate speculative address translation with minimal system modifications. The key idea of Revelator is to have the OS establish a predictable per-page hash-based data placement that the hardware leverages to predict the physical address of requested data. 
Figure~\ref{fig:overview} shows an overview of Revelator's key components and workflow.

\textbf{Key Mechanism.} At the OS level, Revelator employs a \emph{tiered} hash-based allocation policy.
When the OS receives a request to allocate a physical frame for a virtual page~\circled{1}, it computes a bounded sequence of candidate physical page numbers (PPNs), with one candidate per allocation tier (i.e., hash attempt).
For tier $i \in [1,N]$, the OS applies the same hash function $h$ to the VPN and a tier-specific seed $seed_i$, producing $PPN_i = h(VPN, seed_i)$.
The OS attempts to allocate the virtual page to these candidate frames in tier order~\circled{2}. If $PPN_i$ is free, the OS allocates the virtual page to the corresponding physical frame~\circled{3}, establishing a predictable VA-to-PA mapping that the hardware can later recompute using the same VPN and seed.
If $PPN_i$ is occupied, the OS advances to the next tier and computes another deterministic candidate using the same VPN but a different seed.
Because each tier probes an independently chosen candidate frame, a system with memory utilization $u$ gives each candidate an approximately $1-u$ chance of being free, making it increasingly unlikely that all $N$ candidates are occupied, even under high memory utilization. 
If all $N$ candidate physical frames are occupied, the OS falls back to the conventional allocation policy~\circled{4}. We analyze the probability of successful hash-based allocation and its dependence on memory utilization in Section~\ref{sec:os_allocation}.

At the hardware level, when a translation request misses in the TLB hierarchy~\circled{5}, Revelator invokes a lightweight speculation engine that mirrors the OS's tiered hash-based allocation policy. First, the speculation engine uses the same hash function circuit and tier-specific seeds to recompute the candidate $PPNs$ for the miss-causing $VPN$~\circled{6}. Second, it combines each candidate $PPN_i$ with the page offset to form a candidate physical address and issues selected speculative data requests to the memory hierarchy. If a candidate corresponds to the actual physical address~\circled{7}, the speculative data fetch overlaps with the page table walk, and the data can arrive before address translation completes. 
In the example of Figure~\ref{fig:overview}, the OS successfully allocates the virtual page to the candidate frame in tier $i$. However, the hardware does not know a priori which tier the OS used to place the data, so it may also issue \profrevc{extra} speculative fetches for candidates from other tiers, such as tiers 1 and $N$~\circled{8}. In \S\ref{sec:hw_speculation}, we describe how the speculation engine uses a speculation degree filter to reduce this \profrevc{extra} memory traffic by limiting which candidate requests are issued.

\textbf{Key Benefits.} As shown in Table~\ref{tab:contiguity-comparison}, Revelator avoids the limitations of prior contiguity-based mechanisms. 
First, Revelator's prediction accuracy does \emph{not} depend on memory fragmentation but only on memory utilization. This is because Revelator's allocation policy enforces a \emph{per-page} hash-based placement invariant, rather than relying on free contiguous multi-page physical segments. 
Second, as Revelator's effectiveness does not depend on memory fragmentation, it is naturally resilient to allocation interference.
Third, Revelator requires only small OS changes: (i) the OS allocator is modified to support tiered hash-based allocation, and (ii) the OS maintains a small amount of counters per hash function to track the successful allocation rate of each hash function and guide the hardware-based speculation.
Fourth, Revelator requires only small hardware changes: (i) a lightweight speculation engine that computes the same hash function as the OS and (ii) a small number of hardware counters to track the successful allocation rate per hash attempt and guide speculation.

\section{Revelator: Detailed Design}
\label{sec:key_mechanism}

We describe in detail the key components of Revelator's design, including:
(i)~tiered hash-based allocation policy for data pages and its probabilistic properties,
(ii)~rationale behind per-page placement instead of contiguous multi-page placement,
(iii)~extension of tiered hash-based allocation to page table frames,
(iv)~speculation engine and how it dynamically adapts the degree of speculation, and
(v)~how Revelator operates in the presence of large pages, in virtualized environments, and NUMA systems. The extended version of this paper~\cite{revelator-extended} provides a detailed example of Revelator's end-to-end flow of memory allocation and address translation.

\subsection{OS: Tiered Hash-Based Allocation}
\label{sec:os_allocation}

Revelator introduces a tiered hash-based physical page allocation policy that establishes a predictable per-page mapping from virtual page numbers (VPNs) to physical page numbers (PPNs).
For each allocation, the OS computes a bounded sequence of candidate PPNs, with one candidate per allocation tier.
We use the terms \emph{tier} and \emph{hash attempt} interchangeably to refer to one ordered step in this sequence: tier~$i$ uses the same hash function $h$ but a distinct tier-specific seed $seed_i$, producing $\text{PPN}_i = h(\text{VPN}, seed_i)$.
This tiered candidate set narrows the possible physical locations for each VPN, enabling hardware to later recompute the same candidates and speculate on the most likely physical addresses.

\noindent\textbf{Hash-based Allocation Attempt.}
Upon an allocation request, the OS starts at tier~1 and computes $\text{PPN}_1 = h(\text{VPN}, seed_1)$.
It then checks whether $\text{PPN}_1$ is a valid candidate frame and is present in the buddy allocator's free lists.
If the frame is free, the OS removes it from the buddy allocator and maps $\text{VPN}$ to $\text{PPN}_1$.
If the frame is occupied, reserved, or otherwise unavailable, the OS advances to the next tier and computes another deterministic candidate using the same VPN but a different seed.
This process continues for tiers $2,3,\ldots,N$ until either a free candidate PPN is found or all $N$ hash attempts are exhausted. If all $N$ candidate PPNs are unavailable, Revelator falls back to the conventional allocation mechanism to find any available free page $\text{PPN}_{\text{FB}}$.
In this case, the $\text{VPN}$ to $\text{PPN}_{\text{FB}}$ mapping is not predictable.

\noindent\textbf{Probabilistic Analysis.}
Assuming each hash attempt $H_i$ produces candidate PPNs that are uniformly distributed across the physical address space, each hash-based allocation attempt succeeds with a probability of $p = M/P$, where $M$ is the number of free memory pages and $P$ is the total number of memory pages.
Revelator's tiered hash-based allocation mechanism can be modelled as $N$ sequential Bernoulli trials.
Thus, the probability that at least one of the $N$ hash attempts succeeds is $P_S{=} \sum_{i=0}^{N-1} p(1-p)^i {=} 1 - (1-p)^N$.
As $N$ increases, the probability that all candidates are unavailable decreases exponentially as $(1-p)^N$. Our Linux kernel prototype of Revelator validates the exponential decrease (\S\ref{sec:linux-kernel}). Even though increasing $N$ improves the probability of a successful hash-based allocation,
it also increases (i) the number of hash-based allocation attempts the OS may perform during allocation and (ii) the number of candidate physical addresses the hardware needs to fetch before translation completes, potentially increasing allocation latency and memory bandwidth consumption.

\noindent\textbf{Leveraging Tiering-Induced Bias.}
\hs{The probability that allocation is successful at hash attempt $i$ decreases geometrically in $i$: $P(\text{Alloc}_i){=}p(1-p)^{i-1}$ (with $i{=}1$ corresponding to $H_1$).}
\hs{This induces a \emph{tiering bias}: earlier hash attempts produce more allocations than later ones --- $H_1$ produces the most, $H_2$ the next-most, and so on.}
Revelator's speculation mechanism leverages this bias by generating candidate physical addresses in tier order and issuing the most likely speculative fetch first, i.e., the candidate derived from $H_1(VPN)$.
Issuing additional fetches for more candidates such as $H_2(VPN)$ and $H_3(VPN)$ can improve speculation coverage, but it also consumes additional memory bandwidth (\S\ref{sec:heuristic} describes how Revelator dynamically selects how many candidates to fetch).

\noindent\textbf{\hs{Choice of Hash Function}.}
Revelator uses CityHash~\cite{cityhash}, a fast non-cryptographic hash function also used in prior hash-based schemes~\cite{elastic-cuckoo-asplos20,nestedecht,mehtJovanHPCA2023}.
The OS and hardware use the same hash function so that the hardware can deterministically recompute the candidate PPNs generated during allocation.
To generate multiple candidates, Revelator feeds the hash function with the VPN, the process identifier (PID), and a tier-specific counter-based seed: $PPN_i {=} \text{CityHash}(VPN, PID, seed_i)$.\footnote{Using the PID as input to the hash function ensures that the same VPNs across different processes map to different candidate PPNs.}

\subsection{Hash-Based Page Table Allocation}
\label{sec:pt_allocation}

\begin{figure}[!ht]
    \centering
    \includegraphics[width=1.0\linewidth]{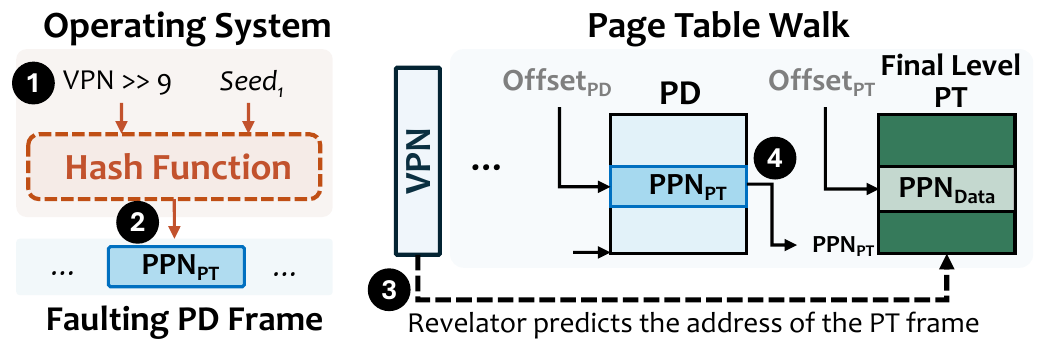}
    \vspace{-3mm}
    \caption{Revelator's hash-based allocation for page table frames.}
    \label{fig:pt_allocation}
    \vspace{-1mm}
\end{figure}

\hs{Revelator also applies hash-based allocation to final-level page table (PT) frames, making the VA-to-PTE mapping predictable and enabling the hardware to speculatively fetch the final PTE at the start of the PTW.}
Unlike data-page allocation, PT-frame allocation uses a single hash attempt rather than multiple tiers \profrevc{(a conservative choice because the potential performance benefit of additional PT-frame hash attempts does not justify the cost of issuing multiple speculative PTE fetches per TLB miss)}.
Figure\hs{~}\ref{fig:pt_allocation} illustrates Revelator's allocation policy for PT frames.
When allocating a final-level PT frame, the OS computes one target PPN using $H_1(VPN {>>} 9)$, where the VPN is right-shifted by 9 because each final-level PT frame contains PTEs for 512 contiguous VPNs~\circled{1}.
If the target PPN is free, the OS allocates the PT frame to that PPN and updates the corresponding page-directory (PD) entry with the selected PT-frame PPN, allowing a conventional PTW to normally reach the PT frame 
~\circled{2}. Otherwise, Revelator falls back to the conventional allocator to find a frame for the PT.
When the hash-based PT-frame allocation succeeds, the MMU can also recompute the same PT frame's PPN from the VPN, combine it with the PTE offset, and speculatively fetch the final PTE at the start of the PTW~\circled{3}.
This overlaps the final PTE fetch with the earlier levels of the PTW, accelerating the walk and reducing address translation latency (\S\ref{sec:hw_speculation}). 

\subsection{Hardware: OS-Guided Speculative Translation}
\label{sec:hw_speculation}

\subsubsection{Speculation Engine}
The speculation engine is part of the memory management unit (MMU) and works in tandem with the OS's tiered hash-based allocation policy to predict the physical page number (PPN) for a given virtual page number (VPN) and fetch the corresponding data before address translation completes.

\noindent\textbf{Speculative PA Calculation.}
\hs{On an L2 TLB miss, the speculation engine extracts the VPN and page offset from the faulting VA.
It then recomputes the same bounded sequence of candidate PPNs that the OS considered during allocation by hashing the VPN with each configured seed.
Each candidate PPN is concatenated with the page offset to form a candidate PA.
\profrevc{For final-level page table frames, the speculation engine similarly hashes $VPN \gg 9$ to predict the PT-frame PPN and combines it with the PTE offset to form a candidate PTE address.}
Because the hardware does not know a priori which hash attempt succeeded during OS allocation, it can generate up to $N$ candidate PAs, one per tier, and issue speculative fetches in the configured tier order.}
The latency of the hash function circuitry is only 2--3 cycles, which is negligible compared to the latency of a PTW ({40--100} cycles).

\subsubsection{Dynamically Adapting Speculation Degree}
\label{sec:heuristic}

Revelator needs to decide how many of the hash-based candidate PPNs to turn into actual speculative memory requests on each L2 TLB miss, which we refer to as the \emph{speculation degree}.
The speculation degree controls a trade-off between address translation coverage and memory system pressure.
A larger degree (e.g., $N{=}4$) covers pages allocated by later tiers and therefore increases the likelihood that at least one speculative fetch targets the correct PPN.
However, for each additional tier, an additional speculative
 memory request needs to be issued, consuming bandwidth and potentially delaying demand accesses.
A smaller degree (e.g., $N{=}1$) avoids unnecessary traffic when most pages are allocated by the first tier, but it misses pages that were allocated by later tiers ($N{>}1$), which become more common as memory utilization increases.
Because the right degree depends on both allocation behavior and available bandwidth, Revelator uses a speculation degree filter to dynamically select how many candidate PAs to fetch.

\noindent\textbf{Memory Utilization:}
As memory utilization increases, the OS is more likely to use secondary tiers ($H_2$, $H_3$, etc.) or the conventional allocator to allocate a page.
In such scenarios, increasing the speculation degree improves speculative hit likelihood.

\noindent\textbf{Memory \hs{bandwidth}:}
When memory bandwidth is abundant, Revelator can speculate on more candidate PAs because the extra requests are less likely to contend and safely improve speculation coverage without hurting demand requests. When bandwidth is scarce, however, speculative requests to candidate PAs that do not match the OS-selected PPN become costly: they occupy memory queues, consume interconnect and DRAM bandwidth, and can delay demand requests.

\begin{figure}[!ht]
    \centering
    \includegraphics[width=1.0\linewidth]{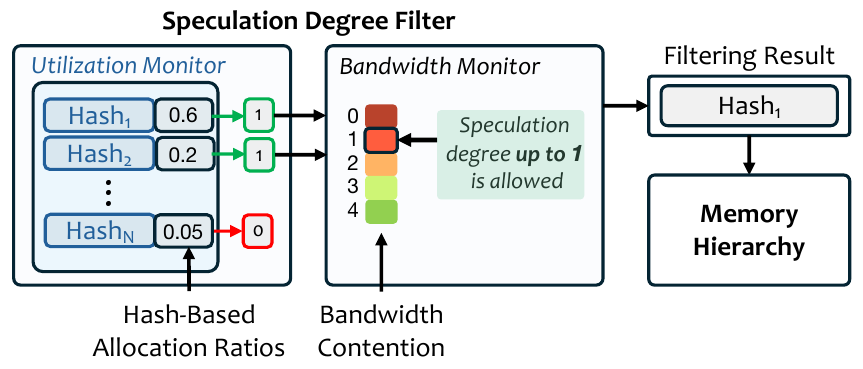}  
    \caption{Speculation degree filter design.}
    \label{fig:spec-engine}
    \vspace{-1mm}
\end{figure}

\noindent\textbf{The Speculation Degree Filter:}
Figure~\ref{fig:spec-engine} shows the design of the speculation degree filter.
\hs{It employs two monitors to dynamically adapt speculation degree.
First, the memory utilization monitor tracks the fraction of all successful allocations performed by each tier.
Second, the bandwidth monitor tracks bandwidth usage.
The speculation degree filter uses the utilization monitor to drop tiers that have allocated fewer pages than a pre-defined threshold.
For the remaining tiers, the filter adaptively adjusts the degree based on the available memory bandwidth.
The final PA candidates are speculatively fetched.}

\subsection{Synergy with Large Pages}
\label{subsec:integration-with-thp}
\label{sec:revelator-thp}

Large pages reduce the overhead of address translation when the OS can find sufficiently large contiguous physical regions, but their benefits degrade in the presence of fragmentation.
Revelator addresses this limitation by making the physical location of 4KB pages
predictable through hash-based placement, while still opportunistically exploiting larger regions when they are available.
This lets Revelator preserve fast translation under fragmentation without giving up the benefits of large pages in low-fragmentation settings.

\noindent\hs{\textbf{Per-Page vs. Multi-Page Hash-based Placement.}}
A natural design question for Revelator is whether hash-based allocation should remain strictly per-page, or instead try to preserve predictable placement for larger multi-page regions, similar in spirit to mechanisms that support multiple page sizes~\cite{guvenilir2020tailored,psomadakis2024elastic,hybridtlbISCA2017,mosaic2017MICRO}.
\hs{Such a design would map $k$ consecutive virtual pages to a hash-derived $k$-page physical region, allowing hardware to speculate at a coarser granularity.}
\hs{The drawback is that predictability now requires contiguity: allocation succeeds only if physical memory contains a \profrevc{sufficiently large contiguous region of free pages}, so success depends on fragmentation, not just total free capacity.}

\begin{figure}[!ht]
    \centering
    \includegraphics[width=1.0\linewidth]{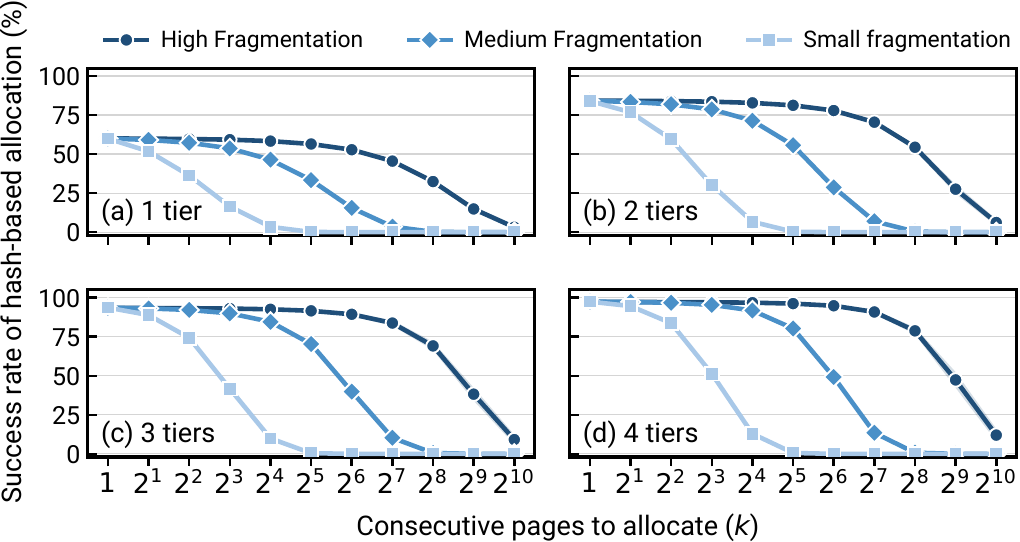}
    \vspace{-2mm}
    \caption{\revCQ{Ratio of successful hash-based allocations for $k$ consecutive pages under different fragmentation patterns and numbers of hash \hs{attempts} \profreva{(i.e., tiers)} in a 40\% utilized memory.}}
    \label{fig:hash-quality}
\end{figure}

Figure~\ref{fig:hash-quality} shows the probability of successfully allocating $k$ consecutive pages using a hash-based allocator with 1-4 allocation tiers (i.e., independent hash \hs{attempts}). This experiment was conducted using the simulation infrastructure discussed in \S\ref{sec:methodology}, by running a custom allocation-heavy microbenchmark.
We use CityHash~\cite{cityhash} as the underlying hash function and during allocation we derive each candidate physical range by hashing the virtual page number at $k$-page granularity (i.e., computing \hs{$\mathit{CityHash}(\mathit{VPN} \gg \log_2 k)$}), where $k$ ranges from 1 to 10.
All experiments use identical memory utilization levels (40\% of memory is \profrevc{allocated}) but different fragmentation levels.

We make two key observations.
First, increasing the number of tiers increases the allocation success probability, because each additional tier provides another independent chance to find a free region.
Second, successful allocations reduce fast as $k$ increases when memory fragmentation is medium/low, even when multiple independent hash \hs{attempts} are available.
In contrast, when $k=1$, \hs{allocation} success depends only on the fraction of free pages and is independent of fragmentation structure.
This observation motivates Revelator's core design choice to enforce a per-page placement ($k=1$) and to exploit 2MB regions only opportunistically, when sufficiently large contiguous regions are available.

\noindent\textbf{Revelator+THP.} Based on this observation, we build Revelator+THP, a variant of Revelator that combines opportunistic 2MB placement with the base 4KB tiered hash-based allocation.
Revelator+THP does not rely on large contiguous regions being available.  It first attempts to exploit 2MB large-page benefits when possible, and then falls back to Revelator's 4KB hash-based placement when large pages are unavailable, ensuring robust performance across different fragmentation levels.
On each data allocation, the OS proceeds as follows:

\noindent\textbf{(1) Hash-based 2MB allocation.}
For a 2MB-aligned virtual region, the allocator first tries a hash-based 2MB page allocation. It hashes the 2MB-aligned virtual address to select a candidate 2MB-aligned physical region and checks whether the entire region is available. If so, the OS allocates that physical region, enabling the hardware to predict data placement at the 2MB granularity.

\noindent\textbf{(2) Conventional THP fallback.}
If hash-based 2MB allocation fails, the OS falls back to the conventional Linux THP allocation scheme~\cite{corbet2011}. THP attempts to find and directly allocate any available free 2MB region. If successful, 
the page is allocated as a 2MB page, and the hardware can predict its placement at the 2MB granularity. 

\noindent\textbf{(3) Hash-based 4KB allocation.}
If no 2MB physical region is free, the allocator falls back to Revelator's  tiered hash-based 4KB allocation. To prevent interfering with future THP opportunities, 4KB hash-based allocation is strictly restricted to physical memory regions not already marked or reserved for 2MB pages. 

\noindent\textbf{(4) Buddy allocator fallback.}
As a last resort, the system falls back to the conventional buddy allocator to place the page in any arbitrary 4KB frame.

\vspace{-1mm}
\subsection{Revelator in Virtualized Environments}
\label{sec:virtualization}

\begin{figure}[!ht]
    \vspace{-2mm}
    \centering
    \includegraphics[width=1.0\linewidth]{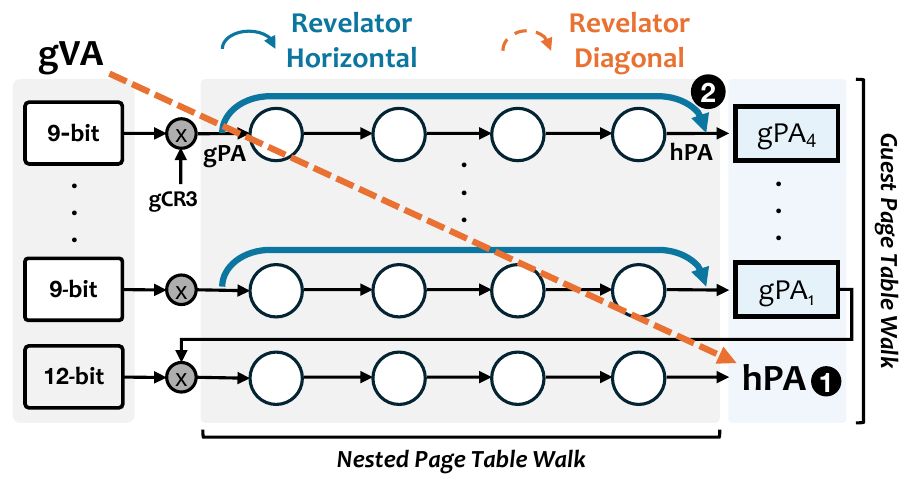}
    \caption{Revelator's workflow in x86-64 virtualized environments that use Nested Paging~\cite{amdnested}.}
    \vspace{-1mm}
    \label{fig:revelator-virt}
\end{figure}

Revelator’s functionality naturally extends to virtualized systems. \profrevc{In this section, we describe Revelator in the context of x86-64 virtualization with Nested Paging~\cite{amdnested}, where address translation requires a two-dimensional page table walk from guest virtual addresses to guest physical addresses and then to host physical addresses.} Our insight is to leverage tiered hash-based allocation and hardware speculation in two \emph{complementary and composable} ways. Figure~\ref{fig:revelator-virt} illustrates the two approaches.

\noindent\textbf{Diagonal speculation.}
First, Revelator can be used to predict the final host physical address (hPA) directly from the guest virtual page number (gVPN), overlapping the \emph{entire} 2D PTW with data fetching (\emph{diagonal speculation},~\circled{1}). 
In this approach, the hypervisor employs tiered hash-based allocation and allocates hPAs using the guest virtual address as the input to the hash function ($Hash(gVA)=hPA$). 
We provide more details on how the hypervisor can implement diagonal speculation without guest cooperation in the extended version of this paper~\cite{revelator-extended}.

\noindent\textbf{Horizontal speculation.}
Second, Revelator can make the guest-physical (gPA) to host-physical (hPA) mapping predictable by having the hypervisor allocate hPAs using Revelator's standard hash-based policy, with the gPA as the hash input (\S\ref{sec:os_allocation}). This enables \emph{horizontal speculation} during nested page table walks. Once the guest PTW identifies the gPA of a guest page-table page, the hardware hashes that gPA to predict the hPA where the page resides. It then speculatively fetches the cache line containing the next-level guest PTE before the nested translation completes~\circled{2}.

    \subsection{Revelator in NUMA Systems}
    \label{sec:NUMA}

    \noindent\textbf{\revCQ{NUMA Placement and Speculation.}}
    \revCQ{In NUMA systems, the OS selects the node from which each page is allocated according to the active memory policy, such as \texttt{Interleave}, \texttt{Local}, \texttt{Bind}, or \texttt{Preferred}~\cite{linux-numa-policy,linux-set-mempolicy,linux-mbind}. This node selection becomes part of the physical address, but for common policies such as \texttt{Local}, \texttt{Bind}, and \texttt{Preferred}, the hardware usually does not know the target node before address translation resolves. This uncertainty makes PA speculation challenging. Speculating across every node wastes bandwidth, while speculating only locally misses optimization opportunities for pages placed remotely due to policy, pressure, or migration. 
    Existing contiguity-based mechanisms (e.g., SpOT~\cite{chloe2020}) face the same NUMA-local fragmentation problem, because they still depend on finding or preserving large contiguous regions within the selected node. If such schemes instead use the memory capacity of a remote node to preserve contiguity, the resulting access latency and interconnect traffic can offset their address translation benefits.
    }

    \noindent\textbf{\revCQ{OS Side: Policy-First Node-Scoped Hashing.}}
\revCQ{Revelator keeps NUMA placement under OS control because the selected NUMA node affects the latency and bandwidth of every later memory access to the page which is often 
more important than reducing address translation latency.   
During allocation, the OS first selects the target node $n^{*}$ according to the active NUMA policy. Revelator then performs tiered hash-based placement only within $n^{*}$'s address range, making intra-node 4KB placement predictable without changing the OS's locality decision. If all intra-node hash attempts fail, Revelator falls back to the conventional allocator within $n^{*}$. This policy-first design preserves OS locality decisions while avoiding the need to find large contiguous physical regions inside the selected node. We focus on \texttt{Local}, \texttt{Bind}, and \texttt{Preferred}, because unlike \texttt{Interleave}, their target node is not directly derivable from the VPN and is therefore challenging for PA speculation.}

\begin{figure}[!ht]
        \centering
        \includegraphics[width=1.0\linewidth]{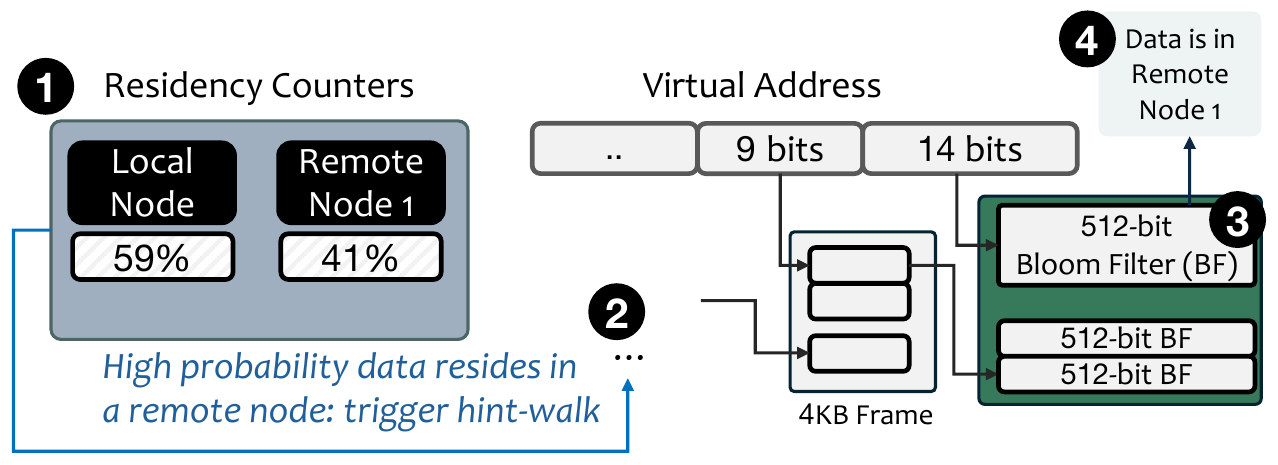}
        \vspace{-2mm}
        \caption{Revelator's workflow in NUMA systems.}
        \label{fig:numa_hints}
        \vspace{-1mm}
    \end{figure}

    \noindent\textbf{\revCQ{Hardware Side (1): Dominant-Driven Speculation.}}
    \revCQ{Figure~\ref{fig:numa_hints} illustrates Revelator's NUMA-aware speculation policy.
    The speculation engine maintains per-NUMA-node counters $C_n$,
    incremented when a translation resolves and data resides in node $n$.
    These counters provide an empirical probability distribution, estimating the likelihood that data resides in node $n$ ($p_n = C_n / \sum_i C_i$~\circled{1}).
    Let $p_{\text{dominant}}$ be the probability of the \emph{dominant} node
    with the highest counter value ($n_{\text{dom}} = \arg\max_n C_n$).
    (Note: the dominant node is not necessarily local).
    If $p_{\text{dominant}} \ge T_{\text{dom}}$ ($T_{\text{dom}}=0.8$),
    Revelator speculates only within the dominant node.
    }

    \noindent\textbf{\revCQ{Hardware Side (2): Hint-Assisted Speculation.}}
    \revCQ{If $p_{\text{dominant}} < T_{\text{dom}}$,
    no single node clearly dominates, increasing the chance that the page resides 
    in a non-dominant node~\circled{2}.
    Without identifying the target node, the engine must either speculate across every node (wasting bandwidth), or forfeit speculation.
    To avoid missing speculation opportunities, Revelator executes two concurrent actions: (i) speculates within the dominant node for likely hits, and (ii) performs a \emph{hint walk}, a parallel lookup in a new OS-managed Bloom-filter table indexed by the VPN, to determine whether the page may reside in a non-dominant node~\circled{3}. The hint walk is not part of the correctness-critical address translation path: it only returns a possible NUMA node for speculation, and the resolved page table translation still determines the correct physical address. If the hint points to a non-dominant node, Revelator issues one additional speculative fetch using the hash-derived candidate address within that node~\circled{4}. We provide more details about this hint structure in the extended version of this work~\cite{revelator-extended}.
    }

\vspace{-1mm}
\section{System Integration}

\vspace{-0.5mm}
\subsection{Security Considerations}
\label{sub:security_considerations}

Revelator \emph{does not} introduce security risks beyond those of a typical cross-page hardware prefetcher~\cite{imp,ainsworth2018event,prodigyHPCA2021,vldp,cooksey2002stateless,ebrahimi2009techniques,amir1998dependence}.
First, Revelator does not allow the processor to use speculatively fetched data along a mis-speculated path, completely preventing Meltdown- or Spectre-style vulnerabilities~\cite{meltdown,schwarz2019netspectre}.
Second, Revelator limits the risk of cache side channels.
An attacker could try to use a cache side-channel attack, such as Prime+Probe~\cite{liu2015last}, to infer \emph{when} a privileged physical address ($PA_{secret}$) is speculatively fetched.
This does not reveal the data stored at $PA_{secret}$, but it could leak coarse-grained information about system activity.
Revelator mitigates this risk in two ways.
First, it uses a secret, per-process hash key, making it difficult for an attacker to compute virtual to physical address collisions that could intentionally steer speculation toward a target address.
Second, once the PTW resolves, hardware removes the microarchitectural trace of incorrect speculative fetches, narrowing the timing window available to a cache side-channel attack.

\begin{figure}[!ht]
    \centering
    \includegraphics[width=1.0\linewidth]{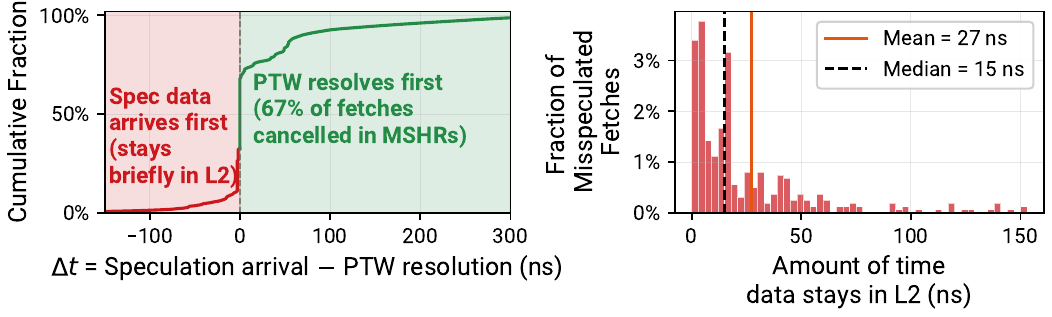}
    \vspace{-4mm}
    \caption{\revF{(Left) Timing of speculative fetch arrival relative to PTW resolution. (Right): Exposure window for cache blocks that arrive in the L2 cache before cleanup.}}
    \label{fig:spec-timing}
    \vspace{-3mm}
\end{figure}

\revF{\noindent\textbf{Speculative Fetch Tracking and Cleanup.}
For each page table walker, the speculation engine records the addresses of outstanding speculative fetches in a small bounded address log.
The number of in-flight fetches is therefore limited by the number of page table walkers and the maximum speculation degree ($N$).
When a speculatively fetched cache line arrives, Revelator installs it only in the private L2 cache, confining any transient microarchitectural footprint to the requesting core and preventing cross-core cache side channels.

When the PTW resolves, the MMU identifies the correct physical address and invalidates any incorrect speculative addresses recorded in the log.
This cleanup uses standard invalidation logic and occurs off the critical path.
Figure~\ref{fig:spec-timing} quantifies why this cleanup significantly narrows the attack window.
The left panel plots the time between speculative data arrival and PTW resolution: for 67\% of mis-speculated fetches, the PTW resolves first, so the request can be cancelled in the MSHR~\cite{kroft1981lockup} before it installs in L2.
Only the remaining fetches arrive early enough to briefly enter the private L2.
The right panel shows that, even for these installed lines, the exposure window is short: the average lifetime is 27~ns before invalidation.
An attacker must therefore probe the victim's private L2 within this narrow interval, making reliable cache-timing attacks substantially harder than against conventional hardware prefetchers that leave longer-lived traces in shared cache structures.
}

\subsection{Integration into the Linux Kernel}
\label{subsec:integration-into-the-Linux-Kernel}
\label{sec:linux-kernel}

We prototype Revelator in Linux 6.10.8~\cite{linux-610} and integrate the tiered hash-based allocation policy with the buddy allocator.
Since Revelator probes at most a bounded number of hash-derived PPN candidates and performs constant-time metadata checks for each candidate, the hash-based allocation path performs $O(1)$ work per allocation.
 We provide extensive details about the Linux integration in the extended version of this paper~\cite{revelator-extended}.

\begin{figure}[!ht]
    \centering
    \includegraphics[width=1.0\linewidth]{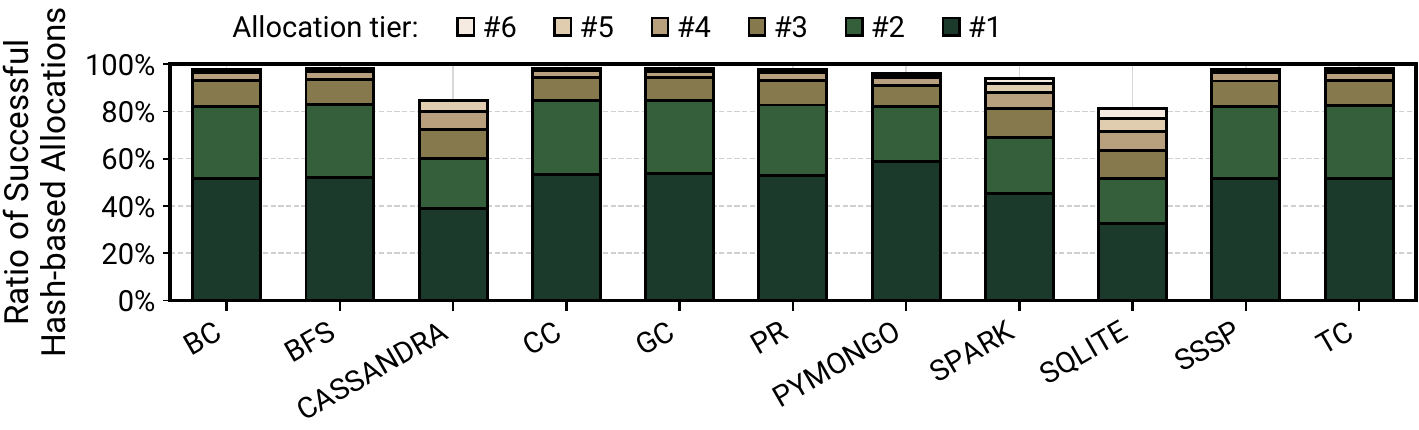}
    \vspace{-5mm}
    \caption{Breakdown of successful allocations using Revelator's tiered hash-based allocation policy measured in a real Linux 6.10.8 prototype.}
    \vspace{-1mm}
    \label{fig:linux_kernel}
\end{figure}

Figure~\ref{fig:linux_kernel} shows the breakdown of Revelator's successful hash-based allocations across 6 allocation tiers, measured on a real system using our prototype implemented on top of Linux 6.10.8~\cite{linux-610}.
During the experiments, a background memory hog continuously allocates and frees pages to keep memory utilization at $\sim$50\%.
To broaden workload diversity beyond the workloads we evaluate in \S\ref{sec:results}, we also evaluate database applications based on Cassandra~\cite{cassandra}, PyMongo~\cite{mongodb}, SQLite~\cite{sqlite}, and Spark~\cite{spark}, which are \emph{not} bottlenecked by address translation.
We make two observations.
First, the first hash tier alone is sufficient to allocate on average 33--59\% of pages, exceeding 50\% for 8 of the 11 evaluated workloads.
Second, using three tiers, Revelator successfully allocates on average 70--85\% of pages in a hash-based manner, exceeding 80\% for 7 of the 11 evaluated workloads.

\begin{figure}[!ht]
    \vspace{-1mm}
    \centering
    \includegraphics[width=1.0\linewidth]{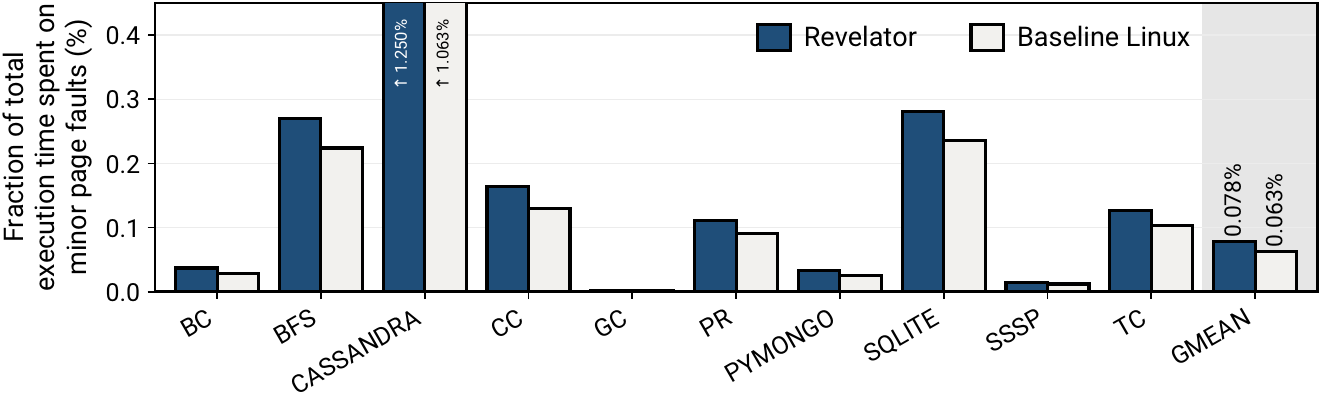}
    \vspace{-5mm}
    \caption{Fraction of total execution time spent on minor page faults in Revelator and baseline Linux across 10 different workloads.}
    \label{fig:overall_execution_time}
    \vspace{-1mm}
\end{figure}

\noindent\textbf{Minor Fault Latency.} 
Figure~\ref{fig:overall_execution_time} shows the fraction of total execution time spent in minor page faults for Revelator and baseline Linux across 10 workloads. We observe that even though baseline Linux's minor fault latency is on average 19\% lower than Revelator's, minor page faults account for only a small fraction of total execution time in both systems (0.06\% for Linux and 0.08\% for Revelator on average). This behavior is expected for long-running translation-intensive workloads, which reach a steady state after their working set is fully mapped and produce no further page faults. Overall, Revelator's additional fault-handling work has a negligible effect on end-to-end application execution time.

\subsection{Area and Power Overheads}
\label{sec:overheads}

We evaluate the area and power overheads introduced by Revelator's hardware. We implement the speculation engine in Chisel~\cite{Bachrachet2012_Chisel} and synthesize it with Yosys~\cite{yosys} using the NanGate 45\,nm open standard-cell library~\cite{nangate45}. The synthesized engine occupies $0.0149\,$mm$^2$ of silicon area and dissipates $14.723\,$mW of static power. This corresponds to only 0.02\% additional area and 0.03\% additional power compared with a high-performance Intel Cascade Lake core~\cite{cascadelake}.

\section{Evaluation Methodology}
\label{sec:methodology}

We evaluate Revelator using Virtuoso~\cite{kanellopoulos2025virtuoso,virtuosogithub}, a validated simulation methodology for accurately modeling the overheads of address translation,
ported on top of Sniper~\cite{sniper,snipergithub}, an event-driven multicore simulator. We extend Virtuoso to model Revelator's OS-level tiered hash-based allocation policy and Sniper to model Revelator's hardware speculation engine. We use CityHash~\cite{cityhash} as the hash function for both the OS and hardware components and charge a 2-cycle latency to produce the hash result. 
Table~\ref{tab:systemconfigs} shows the details of the baseline simulated system configuration. 

\definecolor{SoftPeach}{rgb}{0.937,0.901,0.901}
\begin{table}[h!]
\scriptsize
\centering
\caption{Simulated Baseline System Configurations}
\label{tab:systemconfigs}
\begin{tblr}{
  width = \linewidth,
  colspec = {Q[115]Q[670]},
  cell{0}{1} = {c=2}{0.94\linewidth},
  cell{2}{1} = {r=4}{},
  vline{2} = {1}{-}{0.6pt},
  vline{2} = {2}{-}{0.6pt},
  hline{0-2,6-10} = {-}{},
  hline{3-5} = {2}{},
  vlines
}
\textbf{Core} & 8-way OoO; ROB: 300 entries; x86-64 2.9 GHz \\
\textbf{MMU} & L1 I-TLB: 128-entry, 8-way assoc, 1-cycle  \\
             & L1 D-TLB (4 KB): 64-entry, 4-way assoc, 1-cycle;\newline
             L1 D-TLB (2 MB): 32-entry, 4-way assoc, 1-cycle \\
             & L2 TLB: 2048-entry, 16-way assoc, 12-cycle  \\
             & 3 Page Structure Caches: 4,8,32-entry; 2-cycle  \\
\textbf{L1 Cache} & 64 KB, 8-way, 4-cycle; pLRU;  \\
\textbf{L2 Cache} & 1 MB, 16-way, 12-cycle; pLRU; \\
\textbf{L3 Cache} & 2 MB/core, 16-way assoc, 35-cycle; SRRIP~\cite{srrip}  \\
\textbf{DRAM} & 128GB-DDR4-2400, 4 channels; 2 ranks/channel; 16 banks/rank; \newline tCL = tRCD = tRP = 14.17 ns \\
\end{tblr}

\end{table}

\textbf{Workloads.} Table~\ref{tab:workloads} lists all the workloads used in our evaluation. All evaluated workloads exhibit a high number of page table walks per kilo-instruction (PTWPKI) ($>$5), similar to prior studies~\cite{elastic-cuckoo-asplos20,compendiaISMM2021,midgard,flataAsplos2022, kanellopoulos2023utopia,distributedptMICRO24osang}. Each benchmark is executed for 300M instructions. In single-core systems, we evaluate the 11 most translation-intensive workloads from the GraphBIG~\cite{Lifeng2015}, XSBench~\cite{Tramm2014}, GUPS~\cite{Plimpton2006}, DLRM~\cite{dlmr}, and GenomicsBench~\cite{genomicsbench} suites. In multi-core and NUMA systems, we evaluate 30 mixes from 140 workloads from Google's server suite~\cite{kanev2020workload, google_workload_traces_v2_dynamorio}.

\begin{table}[t!]
  \centering
  \scriptsize
  \caption{Evaluated Workloads}
\vspace{-1mm}
    \begin{tabular}{m{8em}m{19.5em}m{3em}}
    \toprule
    \textbf{Suite} & \textbf{Workload} & \textbf{PTWPKI} \\
    \midrule
    GraphBIG~\cite{Lifeng2015} & Betweenness Centrality (BC), Breadth-first search (BFS), Connected components (CC), Graph coloring (GC), PageRank (PR), Triangle counting (TC), Single-source shortest path (SSSP)   & 10-45 \\
    \midrule
    XSBench~\cite{Tramm2014} & Particle Simulation (XS)     & 12 \\
    \midrule
    GUPS~\cite{Plimpton2006}  & Random-access (RND) & 25 \\
    \midrule
    DLRM~\cite{dlmr}  & Sparse-length sum (DLRM) & 18 \\
    \midrule
    GenomicsBench~\cite{genomicsbench} & k-mer counting (GEN) & 35 \\
    \midrule
    \revCQ{Google Workload Traces~\cite{kanev2020workload, google_workload_traces_v2_dynamorio}} & \revCQ{140 traces from Sierra, Bravo, Tango, Tahoe, Merced, Yankee, Delta, Whiskey} & \revCQ{5-22} \\
    \bottomrule
    \end{tabular}%
  \label{tab:workloads}%
\end{table}%

\begin{figure*}[hb!]
    \vspace{1mm}
    \centering
    \includegraphics[width=1.0\linewidth]{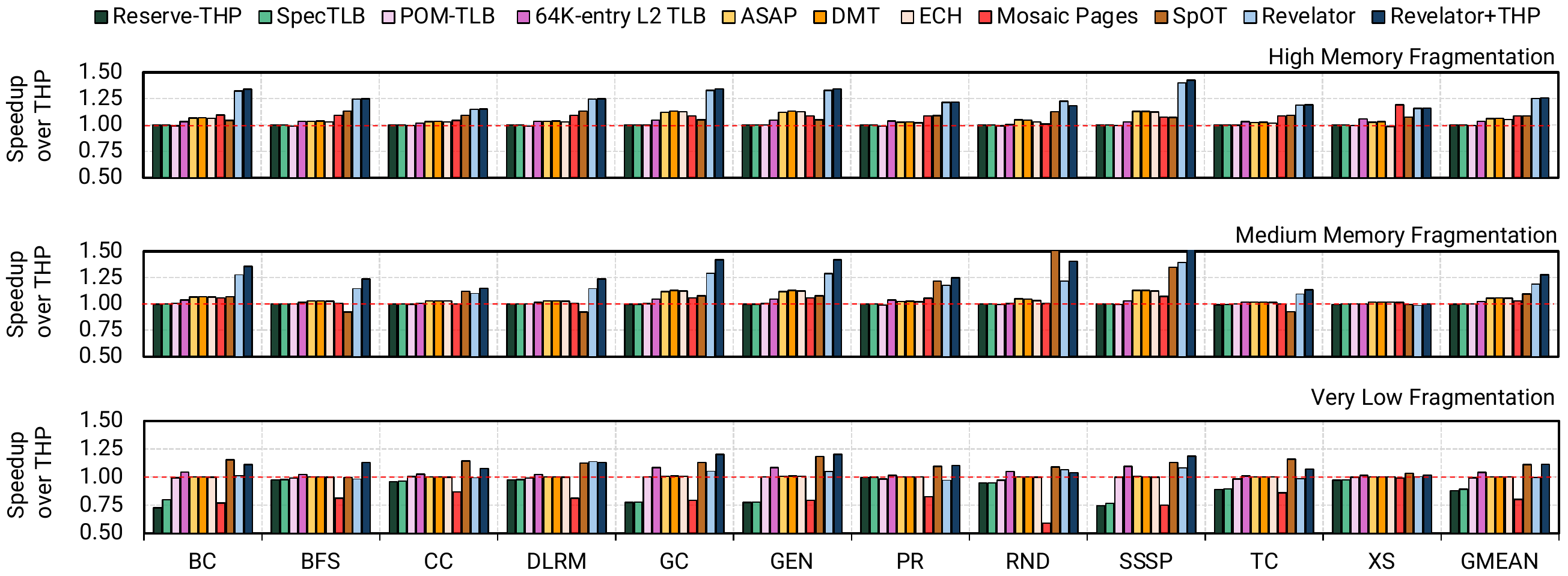}
    \caption{Performance speedup achieved by ReserveTHP~\cite{reserve}, SpecTLB~\cite{spectlbISCA2011}, L2 TLB-64K, POM-TLB~\cite{pomtlbISCA2017}, ASAP~\cite{margaritov2019prefetched}, DMT~\cite{dmtASPLOS2024}, ECH~\cite{elastic-cuckoo-asplos20}, Mosaic Pages~\cite{mosaicpagesASPLOS2023}, SpOT~\cite{chloe2020}, Revelator and Revelator+THP over THP~\cite{reserve} across three  fragmentation levels.}
    \label{fig:results:native::comparison}
\end{figure*}

\textbf{Evaluated Systems in Native Execution.}
We evaluate:
(i) \textbf{\textit{THP}}: Baseline system that utilizes Transparent Huge Pages (4KB/2MB).
(ii) \textbf{\textit{ReserveTHP}~\cite{reserve}}: reserves 2MB regions and promotes them based on the memory utilization of their resident 4KB pages,
(iii) \textbf{\textit{SpecTLB}~\cite{spectlbISCA2011}}: speculates physical addresses based on reserved large page regions, as described in \S\ref{sec:motivation},
(iv) \textbf{\textit{POM-TLB}~\cite{pomtlbISCA2017}}: uses a 16MB software-managed L3 TLB, 
(v) \textbf{\textit{L2 TLB-64K}}: uses a large 64K-entry L2 TLB,
(vi) \textbf{\textit{ASAP}~\cite{margaritov2019prefetched}}: lays out page table data contiguously to enable page table prefetching,
(vii) \textbf{\textit{DMT}~\cite{dmtASPLOS2024}}: establishes a direct mapping from virtual addresses to last-level PT entries,
(viii) \textbf{\textit{ECH}~\cite{elastic-cuckoo-asplos20}}: employs the elastic cuckoo hash-based page table to improve parallelism in page table walks,
(ix) \textbf{\textit{MosaicPages}~\cite{mosaicpagesASPLOS2023}}: employs Iceberg hashing~\cite{iceberghashing} to drastically increase the TLB reach, 
(x) \textbf{\textit{SpOT}~\cite{chloe2020}}: a state-of-the-art speculative address translation scheme, described in detail in \S\ref{sec:motivation},
(xi) \textbf{\textit{Revelator}}: employs tiered hash-based allocation (\S\ref{sec:os_allocation}) and speculation engine (\S\ref{sec:hw_speculation}) with 3 hashes and degree filtering (unless otherwise specified).
(xii) \textbf{\textit{Revelator+THP}}: Revelator combined with THP (\S\ref{sec:revelator-thp}). THP is enabled in L2 TLB-64K, SpOT, POM-TLB, ASAP, DMT, and ECH.

\textbf{Memory Fragmentation \& Utilization.} 
We model fragmentation by controlling how many 2MB pages are available compared to the total possible 2MB pages in the system. 
We use three levels: high, medium, and low fragmentation, which correspond to 10\%, 50\%, and 90\% of the total possible 2MB pages being available, respectively.
These levels match measurements from our lab's 22-node cluster (each node has 64 cores~\cite{kratos20} and 256GB of memory), after approximately one month of uptime, one hour of uptime, and a fresh reboot, respectively.
Fragmentation controls the available contiguity in free memory, whereas memory utilization controls how much memory is already occupied regardless of the data layout. 
We initialize each experiment with 20\% base memory utilization by randomly marking 20\% of the total physical pages as occupied.

\vspace{-1mm}
\section{Evaluation Results}
\label{sec:results}

\subsection{Native Execution Environments}
\label{sec:results:native}
\label{sec:results:comparison}

Fig.~\ref{fig:results:native::comparison} shows the performance provided by Revelator and Revelator+THP compared to 8 state-of-the-art translation schemes (and a 64K-entry L2 TLB) across three fragmentation levels: (top) low, (middle) medium, and (bottom) high fragmentation.

\noindent\textbf{Fast Page Table Walks.} In the low (medium) fragmentation scenario, Revelator outperforms ASAP, DMT, and ECH by 20\% (13\%), 19\% (13\%), and 20\% (13\%), respectively. While effective, these approaches parallelize or reduce the steps of the PTW but the processor still waits for at least one memory access before fetching the data. 
Revelator provides a complementary benefit by overlapping the final data fetch with the PTW, hiding the latency that remains even after the walk itself is accelerated.

\noindent\textbf{Larger or Smarter TLBs.} 
Under low fragmentation, L2 TLB-64K provides a small benefit over THP ($\approx$4\% on average), standalone Revelator roughly breaks even compared to the THP baseline system, and Revelator+THP improves performance by $\approx$12\% by exploiting abundant 2MB contiguity.
This is expected: when 2MB contiguity is abundant, THP already provides high TLB reach while standalone Revelator does not filter 2MB translations in TLBs.
As fragmentation increases, this trend reverses.
Under medium fragmentation, Revelator improves performance by $\approx$19\% over THP, compared to only $\approx$2\% for L2 TLB-64K.
Under high fragmentation, where most memory is backed by 4KB pages, L2 TLB-64K improves performance by only $\approx$3\%, while Revelator improves performance by $\approx$25\%.
Increasing L2 TLB capacity helps only when the additional entries convert misses into hits. This is more likely under low fragmentation, where 2MB pages are abundant and can be captured in TLBs, but becomes less likely as fragmentation increases and more translations require 4KB entries that are not present in the TLB. When a miss still occurs, the processor pays the full PTW cost. Revelator targets this residual cost by starting the data fetch during the PTW, so it can hide latency even for translations that a larger TLB fails to capture.

\noindent\textbf{Speculation-based Schemes.} 
We compare Revelator against SpOT~\cite{chloe2020}, a state-of-the-art contiguity-based speculative address translation technique analyzed in \S\ref{sec:motivation}.
Under low fragmentation, SpOT improves performance by 11\% over THP by exploiting large contiguous mappings, matching the performance of Revelator+THP.
Standalone Revelator roughly breaks even compared to the THP baseline system in this setting because it does not leverage 2MB pages being translated by the TLBs, while SpOT and Revelator+THP do by design.
Under medium fragmentation, SpOT's speedup drops to 9\% as contiguous mappings become harder to preserve.
In contrast, Revelator improves performance by 19\%, outperforming SpOT by 8.5\%, because its hash-based placement does not rely on physical contiguity.
Under high fragmentation, SpOT's speedup drops further to 8\%, while Revelator and Revelator+THP improve performance by 25\% and 26\%, respectively.
This shows that Revelator+THP preserves the benefits of large pages when contiguity exists, while falling back to 4KB hash-based placement and speculation to maintain high performance when contiguity is scarce.

\begin{figure}[h!]
    \centering
    \includegraphics[width=1.0\linewidth]{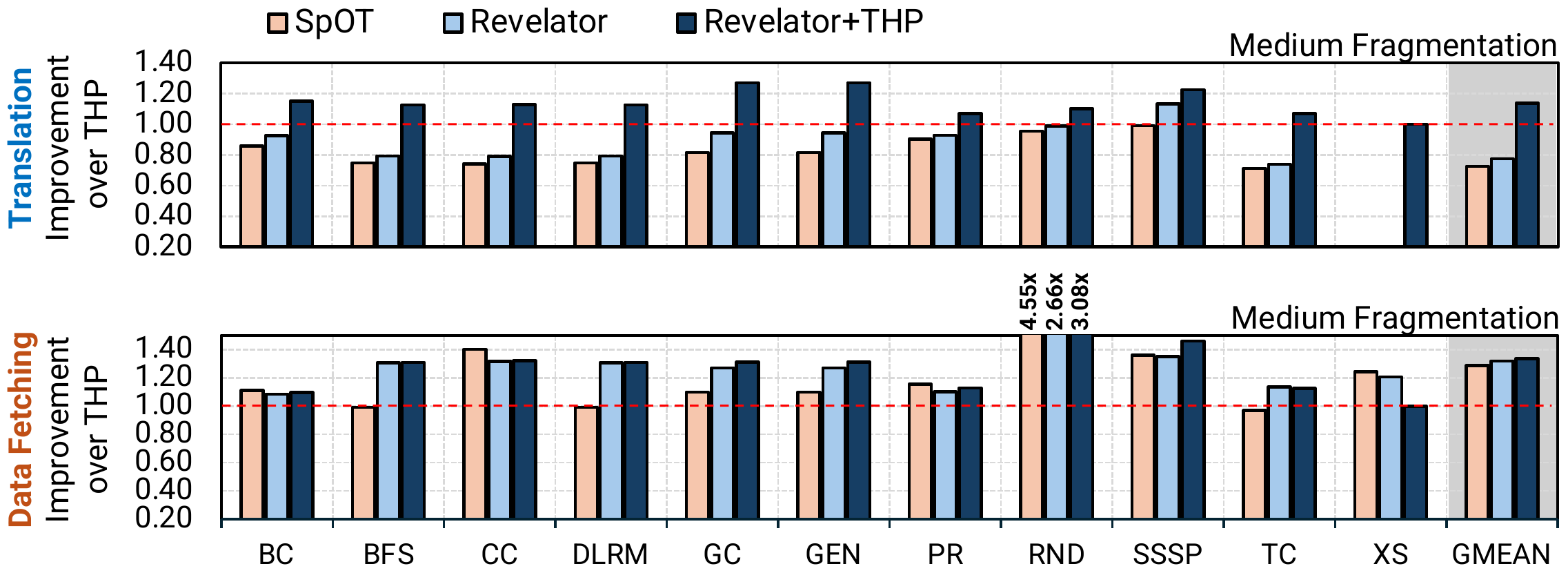}
     \vspace{-3mm}
    \caption{Improvement in address translation latency (top), and data fetch latency (bottom) achieved by SpOT~\cite{chloe2020}, Revelator and Revelator+THP over THP (medium fragmentation) across 11 data-intensive workloads.}
     \vspace{-1mm}
    \label{fig:results:native::explanations}
\end{figure}

\noindent\textbf{Understanding Revelator's Performance.} 
Fig.~\ref{fig:results:native::explanations} breaks down how SpOT (without THP enabled), Revelator, and Revelator+THP affect address translation and data fetch latency in a system with medium fragmentation across 11 data-intensive workloads.
SpOT and standalone Revelator respectively increase address translation latency over THP by 28\% and 23\% on average.
THP remains a strong translation baseline at medium fragmentation: many translations are served by the L2 TLB, and PTWs terminate at the 3rd level due to the presence of 2MB pages.
Revelator+THP reduces address translation latency by 13\% by preserving available 2MB mappings while accelerating PTWs for the remaining 4KB translations.
All three schemes improve data fetch latency by issuing data requests before address translation completes: SpOT, Revelator, and Revelator+THP respectively reduce data fetching by 28\%, 31\%, and 33\% on average.

\begin{figure}[h!]
        \centering
        \includegraphics[width=1.0\linewidth]{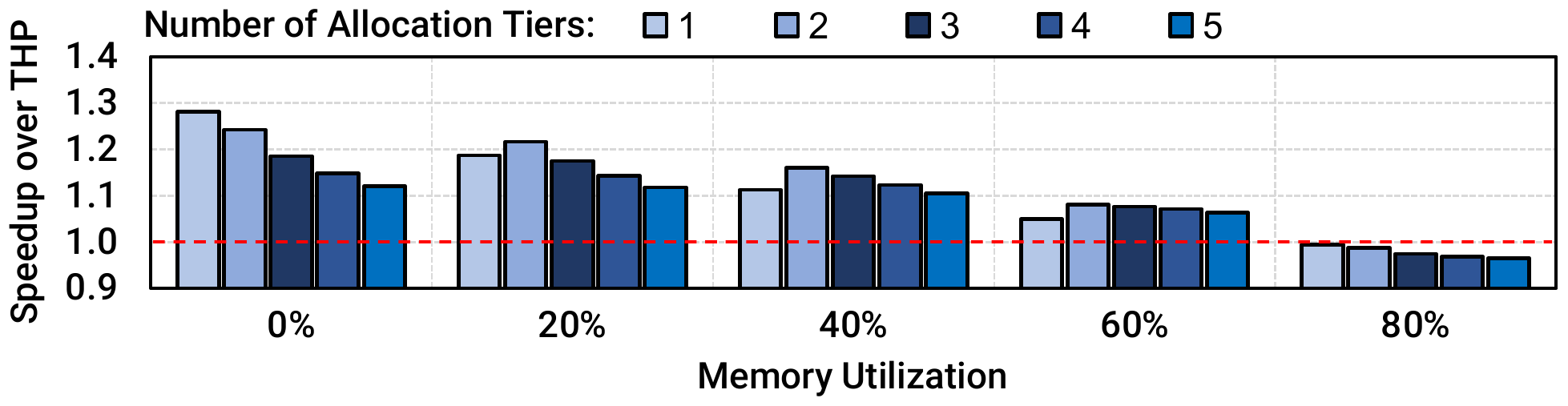}
        \vspace{-5mm}
        \caption{Average speedup over THP achieved by Revelator (medium fragmentation) with up to 5 allocation tiers (i.e., hash attempts) at different memory utilization levels across 11 data-intensive workloads.}
        \label{fig:results:native::fragmentation}
        \vspace{-1mm}
\end{figure}

\noindent\textbf{Impact of Hash Functions and Memory Utilization.} 
Figure~\ref{fig:results:native::fragmentation} shows Revelator's speedup over THP with up to 5 allocation tiers (i.e., hash attempts) as memory utilization increases from 0\% to 80\% in a system with medium fragmentation.
Revelator maintains an 8\% average speedup even under high memory pressure, where 60\% of pages cannot be allocated via a hash function.
The best tier count depends on utilization: one tier performs best at 0\% utilization because the primary target is usually available, while two tiers help at 40\% utilization by increasing the chance of finding a free target.
Using too many tiers adds redundant speculation overhead, causing a 4\% slowdown with four tiers at 80\% utilization. 
These results show that the useful speculation degree should track how many allocation tiers are likely to succeed at a given memory utilization: each additional tier creates another hash-derived candidate address that hardware may speculatively fetch, but fetching candidates from unlikely tiers wastes bandwidth. This motivates the speculation degree filter described in \S\ref{sec:hw_speculation}, which dynamically selects how many tier candidates Revelator should issue as speculative requests.

\noindent\textbf{Additional Sensitivity Studies.}
We further evaluate three sources of Revelator's behavior: (1) the contribution of speculating for page table entries versus the final data fetch, (2) the speculation degree filter's ability to reduce bandwidth overhead, and (3) the cache pollution caused by mis-speculation.
These studies show that Revelator's speedup primarily comes from hiding the final data fetch, that degree filtering reduces wasted fetches without sacrificing performance, and that cache blocks of mis-speculated data rarely evict useful data from the L2 cache.
We provide the detailed analysis for all three studies in the extended version of this paper~\cite{revelator-extended}.

\begin{figure}[h!]
    \centering
    \includegraphics[width=1.0\linewidth]{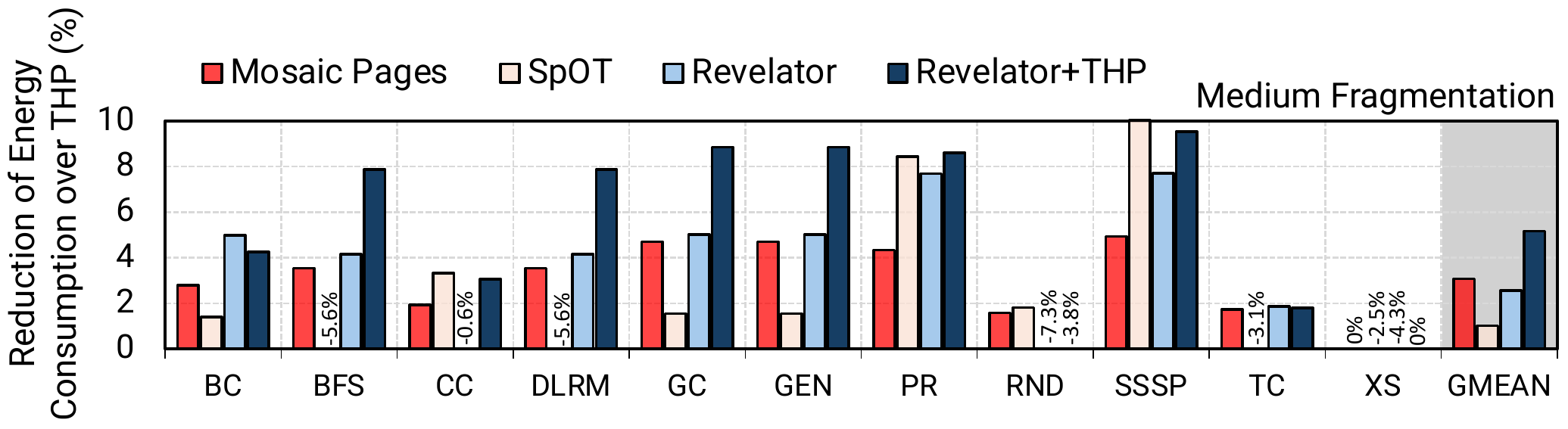}
    \vspace{-5mm}
    \caption{Reduction in energy consumption achieved by Mosaic Pages~\cite{mosaicpagesASPLOS2023}, SpOT~\cite{chloe2020}, Revelator and Revelator+THP over THP in a system with medium fragmentation across 11 data-intensive workloads.}
    \vspace{-1mm}
    \label{fig:results:native::energy}
\end{figure}

\noindent\textbf{Impact on Energy Consumption.} Using McPAT~\cite{mcpat}, in Fig.~\ref{fig:results:native::energy} we evaluate the energy consumption of Mosaic Pages, SpOT, Revelator, and Revelator+THP compared to THP under medium fragmentation across 11 data-intensive workloads. We observe that Revelator+THP reduces energy consumption by 5.5\% over THP and 2.2\%, 4.3\%, and 2.7\% over Mosaic Pages, SpOT, and standalone Revelator, respectively. We provide a comprehensive energy analysis in the extended version of the paper~\cite{revelator-extended}.

\noindent\textbf{Revelator in Virtualized Environments.} We evaluate Revelator in virtualized environments and compare Nested Paging+THP (NP-THP) against DMT~\cite{dmtASPLOS2024} and three Revelator configurations: \texttt{Horizontal} (\S\ref{sec:virtualization}), \texttt{Diagonal} (\S\ref{sec:virtualization}) and \texttt{Full} (\S\ref{sec:virtualization}), which combines both forms of speculation.
\texttt{Horizontal} provides little benefit because the nested translations are often retrieved by the nested TLB, while \texttt{Diagonal} improves performance by 11\% over NP-THP and 6\% over DMT.
\texttt{Full} combines both forms of speculation
and reaches 13.6\% average speedup over NP-THP.
We provide extended analysis on virtualized environments in the extended version of the paper~\cite{revelator-extended}.

\begin{figure}[h!]
    \centering
    \includegraphics[width=1.0\linewidth]{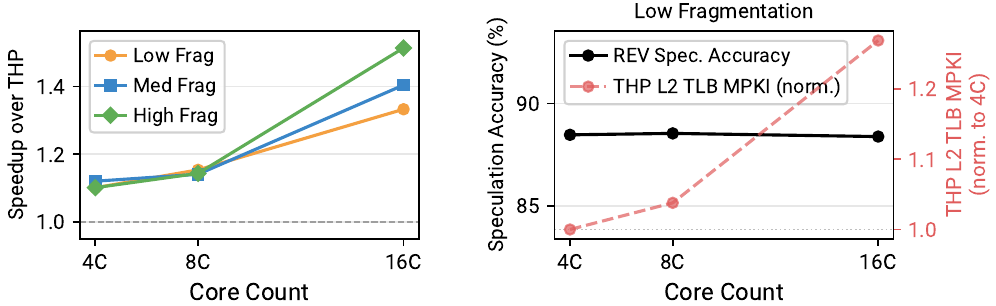}
     \vspace{-5mm}
    \caption{\revCQ{(Left) Average speedup achieved by Revelator over THP under low/medium/high fragmentation; (right) L2 TLB MPKI increase for THP versus speculation accuracy for Revelator under low fragmentation across 4/8/16-core systems.}}
     \vspace{-1mm}
    \label{fig:multicore-scalability}
\end{figure}

\noindent\textbf{Revelator in Multicore Systems.} We evaluate Revelator and THP in 4-, 8- and 16-core configurations across 30 mixes of Google server traces (we compare against SpOT only in the multicore-NUMA evaluation to reduce the total volume of experiments).
Fig.~\ref{fig:multicore-scalability} (left) shows that Revelator consistently outperforms THP as core count increases, providing average speedups ranging from 1.08$\times$ under low fragmentation on 4 cores to 1.50$\times$ under high fragmentation on 16 cores. The benefit grows with core count because more applications compete for 2MB pages, causing THP to fall back to 4KB pages more often while Revelator does not rely on contiguity and continues to hide translation overhead.
Fig.~\ref{fig:multicore-scalability} (right) supports this trend: THP's L2 TLB MPKI increases by more than 20\% from 4 to 16 cores, while Revelator's speculation accuracy remains stable at 87-88\%.

\begin{figure}[h!]
    \centering
    \includegraphics[width=1.0\linewidth]{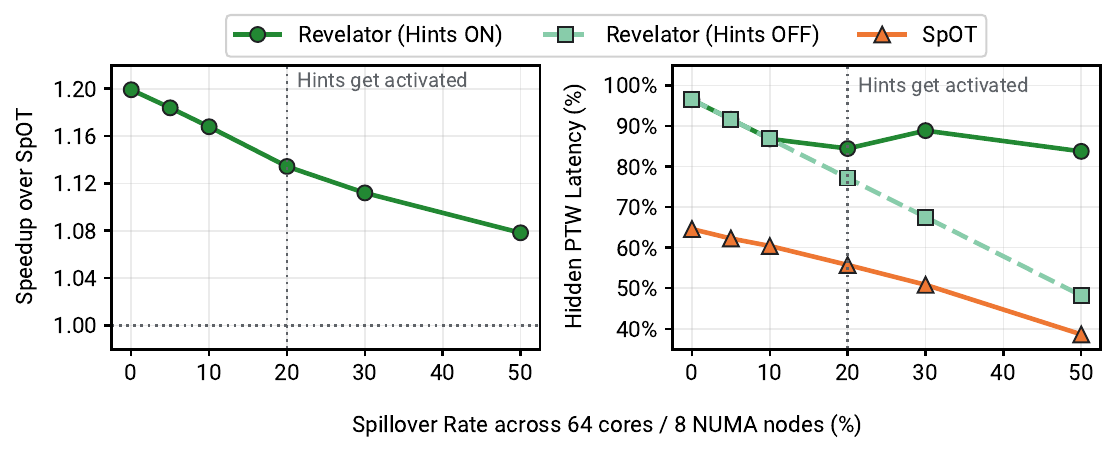}
     \vspace{-5mm}
    \caption{Average speedup and hidden PTW latency achieved by Revelator with Hints ON/OFF over SpOT in an 8-node/64-core NUMA system at different spillover rates across 30 mixes of Google server traces.}
     \vspace{-1mm}
    \label{fig:results:numa}
\end{figure}

\noindent\textbf{Revelator in NUMA Systems.}
We evaluate Revelator in an 8-node/64-core NUMA system across 30 mixes of Google server traces while varying the spillover rate from 0\% to 50\%, where spillover is the fraction of an application's pages allocated outside its dominant NUMA node. The OS policy allocates pages on the dominant node (\texttt{preferred}) until it reaches a per-node memory utilization threshold, at which point it starts allocating on other nodes, increasing spillover.
Fig.~\ref{fig:results:numa} shows Revelator's performance speedup over SpOT~\cite{chloe2020} and the fraction of PTW latency hidden by Revelator (Hints ON/OFF \S\ref{sec:NUMA}) and SpOT.
Revelator consistently outperforms SpOT, retaining an average 1.08$\times$ speedup even at 50\% spillover (up to 1.20$\times$ speedup at 0\% spillover), where many pages reside outside the dominant node.
Bloom-filter-based hints sustain this benefit by directing speculation to likely remote nodes, allowing Revelator to hide 85--95\% of PTW latency across spillover rates.
We provide a detailed NUMA analysis, including additional system configurations and bandwidth overheads, in the extended version of the paper~\cite{revelator-extended}.

\section{Related Work}
\label{sec::related_work}

To our knowledge, \system is the first work that employs hashing to enable highly accurate speculative address translation. 
\S\ref{sec:results:comparison} quantitatively compares \system with 8 state-of-the-art mechanisms that mitigate translation overheads via faster page table walks~\cite{margaritov2019prefetched,elastic-cuckoo-asplos20,dmtASPLOS2024}, larger TLB reach~\cite{reserve,pomtlbISCA2017,mosaicpagesASPLOS2023} and contiguity-based speculative address translation~\cite{spectlbISCA2011,chloe2020}.
This section places \system in the context of prior work that leverages hashing, large pages, contiguity, and TLB optimizations to mitigate address translation overheads.

\noindent\textbf{Hash-based Address Translation \& Page Tables.}
Several works~\cite{elastic-cuckoo-asplos20,distributedptMICRO24osang,hash_dont_cache,nearmemoryPact17,mehtJovanHPCA2023,nestedecht,mosaicpagesASPLOS2023,smith,kanellopoulos2023utopia} leverage hashing to make translation structures smaller, faster, or increase parallelism.
 EMT~\cite{chai2025emt} provides an OS framework for designing such new address translation architectures, which could facilitate \system's integration in future systems.
 Revelator can be tailored to work in tandem with schemes~\cite{kanellopoulos2023utopia,mosaicpagesASPLOS2023} that employ hash-based virtual-to-physical mappings to perform fast address translation (see extended version~\cite{revelator-extended}), and is fully orthogonal to hash-based page tables: Revelator hides the latency of PTWs while hash-based page tables make PTWs faster.

\noindent\textbf{Using Large Pages.}
A large body of works leverages large pages to mitigate address translation overheads\VMlargepages. However, they require contiguous physical memory, which can be scarce in fragmented systems~\cite{mansi2024characterizingphysicalmemoryfragmentation,cbmm,contiguitas2023}.
Prior work improves large-page availability despite fragmentation, by allowing holes inside 2MB pages~\cite{park2020perforated}, by supporting more flexible page sizes~\cite{guvenilir2020tailored,psomadakis2024elastic} or by selectively allocating large pages based on runtime conditions~\cite{ingensOSDI2016,panwar2019hawkeye,tridentMICRO2021}.
In \S\ref{sec:revelator-thp} and \S\ref{sec:results:comparison}, we show that \system works synergistically with large pages. Revelator+THP preserves the benefits of 2MB pages when they are available, while \system continues to accelerate address translation for 4KB pages when fragmentation prevents large-page allocation. 

\noindent\textbf{Leveraging Contiguity.}
Another class of works\VMcontiguity~proposes increasing the effective address translation reach by exploiting contiguous virtual-to-physical mappings.
PTE coalescing mechanisms such as CoLT~\cite{vm6,pham2014} opportunistically merge adjacent contiguous translations into larger TLB entries, while Hybrid TLB Coalescing~\cite{hybridtlbISCA2017} encodes contiguity in PTEs to adapt coalescing granularity under fragmented allocations.
These mechanisms are effective when contiguous ranges can be formed, but their benefits diminish as physical memory becomes fragmented.
As we show in \S\ref{sec:results:comparison}, \system's hash-based placement enables hardware to predict data and PTE locations without depending on the availability of contiguous physical ranges, allowing \system to remain effective under high fragmentation.

\noindent\textbf{Large TLBs, TLB Management \& TLB Prefetching.}
A broad class of works reduces the frequency of PTWs by increasing the TLB reach or by fetching translations before they are demanded\VMtlball.
Large or smarter TLB organizations\VMtlbopts{}, additional TLB levels\VMtlblthree{}, and software-managed TLBs\VMsoftwareTLB{} increase the number of translations that can be cached on-chip.
TLB prefetchers predict future translations and insert them ahead of use\VMtlbprefetching{}, while replacement policies\VMtlbreplacementpolicy{} and cache-backed TLB storage\VMtlbincache{} keep useful translations resident for longer.
These mechanisms are orthogonal to \system: they reduce how often PTWs occur, whereas \system hides the latency of PTWs that TLBs cannot eliminate. 
\section{Conclusion}

This paper introduces Revelator, a hardware-OS cooperative mechanism that leverages hash-based memory allocation to enable highly accurate, low-cost speculative address translation.
Revelator uses tiered hash-based allocation for both program data and last-level PTEs, enabling lightweight hardware to predict where the data and the final PTE are likely to reside after an L2 TLB miss.
By issuing these requests before conventional translation completes, Revelator both hides the address translation latency and accelerates PTWs.
Across single-core, multi-core, virtualized, and NUMA systems, we show that Revelator delivers substantial performance and energy benefits with very small hardware and OS modifications. 
Beyond conventional CPU-based systems, we hope that Revelator's OS-guided predictability provides a foundation for improving address translation in heterogeneous platforms that combine CPUs, GPUs, accelerators and other emerging memory technologies and computing paradigms.
Revelator is freely available at \url{https://github.com/CMU-SAFARI/Virtuoso}.
\section{Acknowledgments}

We thank the anonymous reviewers of MICRO 2025, ASPLOS 2026 and ISCA 2026 for their feedback.
We thank the SAFARI Research Group members for their constructive feedback and for providing a stimulating intellectual and scholarly environment. 
We acknowledge the generous gift funding provided by our industrial partners (especially Google, Huawei, Intel, Microsoft), which has been instrumental in enabling the research we have been conducting on memory systems.

\bibliographystyle{unsrt}
\setlength{\bibsep}{0.25ex}
\bibliography{refs}

\end{document}